\documentclass[sigplan,nonacm]{acmart}

\AtBeginDocument{%
  }

\setcopyright{acmlicensed}
\copyrightyear{2018}
\acmYear{2018}
\acmDOI{XXXXXXX.XXXXXXX}

\acmConference[Conference acronym 'XX]{Make sure to enter the correct
  conference title from your rights confirmation email}{June 03--05,
  2018}{Woodstock, NY}

\acmISBN{978-1-4503-XXXX-X/2018/06}

\usepackage[english]{babel}
\usepackage[utf8]{inputenc}
\usepackage{algorithm}
\usepackage[noend]{algpseudocode}
\usepackage{pifont}
\usepackage{array}
\usepackage{minted}
\usepackage{tikz}
\usetikzlibrary{quantikz2}
\usetikzlibrary{arrows.meta, positioning}

\usepackage[most]{tcolorbox}

\newcommand{\markyes}{\ding{51}}%
\begin{document}

\title{A Closeness Centrality-based Circuit Partitioner for Quantum Simulations}

\author{Doru Thom Popovici\textsuperscript{1}, Harlin Lee\textsuperscript{2}, Mauro Del Ben\textsuperscript{1}, 
Naoki Yoshioka\textsuperscript{3}, Nobuyasu Ito\textsuperscript{3}, Katherine Klymko\textsuperscript{1}, Daan Camps\textsuperscript{1}, Anastasiia Butko\textsuperscript{1}}
\affiliation{
  \textsuperscript{1}\institution{Lawrence Berkeley National Lab}
  \country{USA}
}

\affiliation{
  \textsuperscript{2}\institution{University of North Carolina Chapel Hill}
  \country{USA}
}
\affiliation{
  \textsuperscript{3}\institution{RIKEN RCCS}
  \country{Japan}
}

\renewcommand{\shortauthors}{TBD et al.}

\begin{abstract}
Simulating quantum circuits (QC) on high-performance computing (HPC) systems has become an essential method to benchmark algorithms and probe the potential of large-scale quantum computation despite the limitations of current quantum hardware. However, these simulations often require large amounts of resources, necessitating the use of large clusters with thousands of compute nodes and large memory footprints.  In this work, we introduce an end-to-end framework that provides an efficient partitioning scheme for large-scale QCs alongside a flexible code generator to offer a portable solution that minimizes data movement between compute nodes.  By formulating the distribution of quantum states and circuits as a graph problem, we apply closeness centrality to assess gate importance and design a fast, scalable partitioning method. The resulting partitions are compiled into highly optimized codes that run seamlessly on a wide range of supercomputers, providing critical insights into the performance and scalability of quantum algorithm simulations.
\end{abstract}




\maketitle

\section{Introduction}
Quantum computers are an emerging class of computational systems with the potential to address problems that may be intractable for classical systems. Their anticipated applications include cryptography~\cite{nielsen2010quantum, scarani2009security}, drug discovery~\cite{santagati2024drug}, optimization~\cite{farhi2014quantum}, artificial intelligence~\cite{alexeev2024artificial}, materials science~\cite{bauer2020quantum}, 
high-energy physics simulations~\cite{PRXQuantum.5.037001}, and fundamental studies in quantum chemistry~\cite{mcardle2020quantum}. Several distinct quantum hardware technologies, such as superconducting qubits~\cite{kjaergaard2020superconducting}, trapped ions~\cite{bruzewicz2019trapped}, neutral atoms~\cite{saffman2016quantum}, and photonics~\cite{slussarenko2019photonic}, are under active development, each presenting its own combination of technical challenges, including qubit coherence, gate fidelity, control precision, and scalability. Current devices remain in the Noisy Intermediate-Scale Quantum (NISQ) regime~\cite{preskill2018quantum}, where qubits are highly sensitive to environmental disturbances, leading to frequent errors. Combined with the relatively small number of qubits available, this limits the execution of large and complex quantum circuits and reinforces the need for quantum simulators. Despite these hardware constraints, the development of quantum algorithms that exploit superposition, entanglement, and interference continues to accelerate. Validating these algorithms typically requires large-scale quantum simulations on classical high-performance computing systems, which can consume thousands of compute nodes. Such simulators are essential for assessing algorithmic performance, guiding hardware design, and advancing the field until fault-tolerant quantum computers become practical~\cite{10.1145/3126908.3126947}.

Building on this need, the choice of simulation strategy has a direct impact on scalability, accuracy, and resource requirements. Among the available methods,
Schrödinger-style simulators~\cite{2020arXiv200800216F}, which maintain the full quantum state vector in memory, offer high simulation accuracy but require large amounts of memory. For example, running a quantum circuit of $40$ qubits requires up to $16$ terabytes of data. To execute simulations with large numbers of qubits, the state vector must be split across multiple nodes and efficient data movement techniques must be deployed. 

Over the past years, several approaches have been developed to address the challenge of minimizing data movement when distributing the large state vector. Noteworthy are two main directions: 1) \textbf{graph-based simulators}, which represent quantum algorithms as computational graphs and apply graph partitioning techniques to optimize state-vector distribution~\cite{graph_partitioning, pastor2025community};  2) \textbf{compiler-driven frameworks}, such as Atlas~\cite{10.1109/SC41406.2024.00087} and Quartz~\cite{xu2022quartz}, which use Satisfiability Modulo Theories (SMT) solvers to synthesize distributed quantum programs. While both approaches are effective, they come with limitations. The graph-based approaches often overlook hardware-specific optimizations, while SMT-based methods struggle to scale with increasing quantum algorithm complexity, e.g. finding solutions using solvers like Z3~\cite{de2008z3} requires significant computation time. Moreover, most of the implementations focus on rigid solutions that offer support for specific platforms and vendors, making it difficult for researchers to use the proposed frameworks on the resources available to them. These limitations motivate the need for a more adaptable and performance-oriented approach, one that can account for hardware diversity, exploit memory hierarchies, and reduce communication overhead without being tied to a specific platform.

To address these shortcomings, we propose a flexible framework for transpiling quantum circuits into efficient,  high-performance implementations on both CPU and GPU platforms. The proposed framework represents quantum circuits as graphs and uses closeness centrality to identify critical nodes within the graphs. The framework uses this information to partition the circuit to not only minimize data movement across compute nodes, but also improve memory accesses within each compute node. Since most modern systems feature multiple levels of memory with varying latencies and bandwidths, the framework applies the graph-based analysis recursively for each specified memory level. Finally, the framework interprets the produced partitions and generates the necessary computation and data movement tasks tailored for any CPU and GPU platforms. 

We demonstrated the effectiveness of developed framework on three of today’s leading high-performance computing systems, i.e. Perlmutter at NERSC~\cite{perlmutter2022nersc}, Frontier at Oak Ridge National Laboratory~\cite{frontier2022ornl}, and Fugaku at RIKEN\cite{fugaku2020riken}, covering diverse architectures that include GPU-accelerated platforms as well as CPU-only supercomputers. Our evaluation covers a diverse set of quantum circuit benchmarks with varying qubit counts and gate complexities. For each benchmark, we analyze both the time required to generate partitions and the overall execution time. Results are compared directly against state-of-the-art graph-based and compiler-based frameworks introduced earlier, highlighting its advantages in scalability, portability, and performance. 

This paper makes the following contributions: \begin{itemize} 
\item \textbf{Novel graph-based optimization:} We introduce a method that applies closeness centrality to guide partitioning of quantum circuits, reducing communication costs and improving execution efficiency on classical resources. 
\item \textbf{Hierarchical memory-aware strategy:} We design an optimization approach that aligns with modern multi-level memory hierarchies for both CPU and GPU platforms.
\item \textbf{Flexible and portable framework:} We develop a framework capable of generating efficient distributed implementations for different CPU and GPU systems.
\item \textbf{Comprehensive evaluation:} We demonstrate the effectiveness of our approach on leading HPC systems (Perlmutter, Frontier, and Fugaku), showing simulation performance that is competitive with, and in some cases surpasses, state-of-the-art tools.
\end{itemize}

The rest of the paper is organized as follows. Section~\ref{sec:motivation} discusses the motivation for this work, analyzing state-of-the-art approaches and summarizing their main features. Section~\ref{sec:background} provides a brief introduction to quantum circuits and quantum gates, explaining how they are represented numerically on classical machines. Section~\ref{sec:method} details the proposed method and its implementation. Section~\ref{sec:results} presents the evaluation results. Finally, Section~\ref{sec:conclusions} concludes the paper and outlines directions for future work.

\section{Motivation}
\label{sec:motivation}
First, many frameworks and simulators provide \textit{rigid implementations} tailored to a narrow set of hardware platforms. Extending them to different systems often proves cumbersome and error-prone. Second, most rely on \textit{hand-written kernels} or in some cases \textit{black-box implementations}, for applying quantum gates to the state vector or for fusing gates to improve data locality. Yet the structure of quantum circuits varies significantly across algorithms, making it infeasible to manually implement all possible combinations of fused gates.
Third, numerous frameworks require \textit{expensive analysis} steps when mapping circuits to classical hardware. Some employ computationally heavy graph-partitioning techniques, while others encode each gate as a set of inequalities to be solved by an SMT solver. As circuit depth grows, the cost of such analysis increases sharply. Moreover, analysis and decision times are highly sensitive to hardware characteristics, such as memory hierarchy. Table~\ref{tab:related-work} compares state-of-the-art frameworks for quantum circuit simulation on classical hardware. 

\begin{footnotesize}
\noindent
\begin{table}[t]
\caption{\label{tab:related-work}
Summary of relevant work for Quantum Circuit (QC) simulations and their features: Supported Platform (SP) - CPU (C) or GPU (G); Multi Vendor (MV) - Yes (\markyes) or No; Local Computation (LC) - External Kernels/Libraries (E), Hand-Code Kernels (H) or Code Generated Kernels (C); Data Distribution (DD) - MPI (M), PGAS (P), None (-) or Other (O); Circuit Partitioning (CP) - SMT solvers (S), Greedy Graph Partitioning (G), None (-) or Not Specified (?).
}
\begin{tabular}
{p{1.2in}|p{0.2in}|p{0.19in}|p{0.17in}|p{0.25in}|p{0.21in}}
\hline
Framework
& SP & MV & LC & DD & CP
\\
\hline
qHiPSTER~\cite{2016arXiv160107195S}
& C
& 
& H
& M
& ?
\\
QuEST \cite{jones_quest_2019}
& C, G
& \markyes
& H
& M
& ?
\\
Qiskit~\cite{qiskit2024}+cuQuantum~\cite{2023arXiv230801999B}
& G
& 
& E
& M
& G
\\
SV-Sim~\cite{10.1145/3458817.3476169}
& C, G
& \markyes
& H
& P
& G
\\
Atlas~\cite{10.1109/SC41406.2024.00087} 
& G
&
& E
& O
& S
\\
LFQS~\cite{lfsq} 
& C, G
& \markyes{}
& C
& -
& -
\\
{\bf This work}
& C, G
& \markyes
& C
& M, O
& G
\\
\hline
\end{tabular}
\vspace{-2ex}
\end{table}
\end{footnotesize}

\begin{figure*}[t]
    \centering
    \includegraphics[width=1\linewidth]{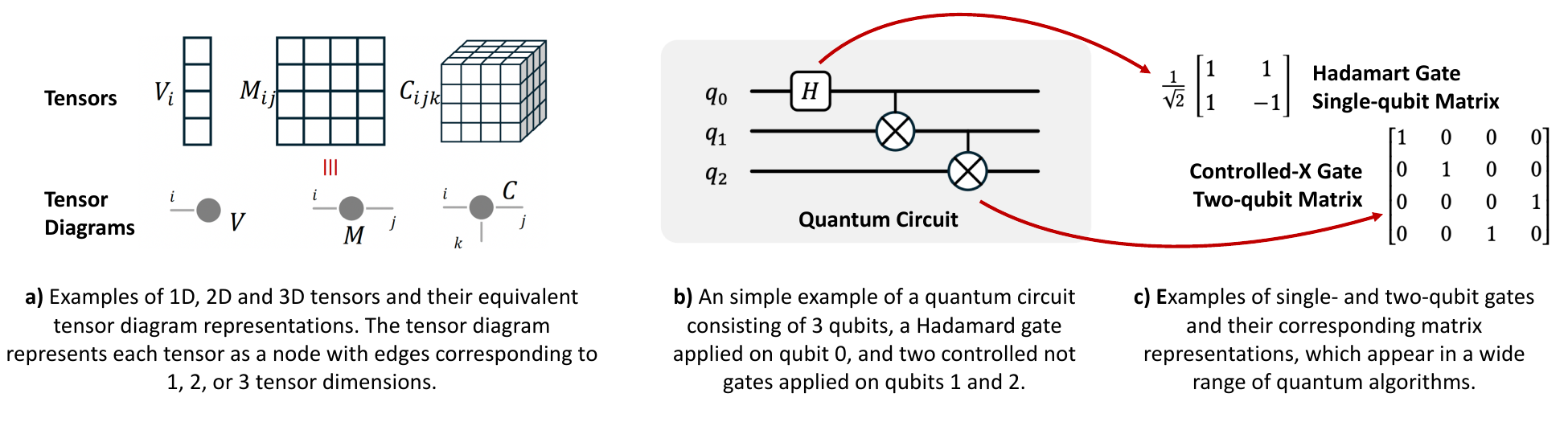}
    \caption{Quantum Circuits Building Blocks.}
    \label{fig:basics}
\end{figure*}

In this work, we address these shortcomings by developing an \textit{end-to-end framework} that enables (1) \textit{fast analysis of quantum circuits} and (2) \textit{automatic, efficient code generation for both CPU and GPU platforms}. While partitioning a graph is an NP-hard problem, we adopt a \textit{greedy strategy based on closeness centrality}, which is computationally cheap and provides a practical measure of computational importance. We further introduce a lightweight code generator, inspired by the TACO compiler~\cite{kjolstad2017tensor}, to automatically produce code for heterogeneous platforms. No parts of the computation or data movement are hand-coded; everything is generated automatically. Finally, we will show that the developed approach can not only produce competitive implementations for NVIDIA platforms, but it can also target AMD GPUs and CPU-based systems.

\section{Background}
\label{sec:background}

In this section, we provide a brief introduction to quantum circuits and quantum gates, with a focus on their standard matrix–vector representation, which is commonly used to model gate operations and circuit execution on classical hardware. This representation supports most quantum circuit simulators and serves as the foundation for our approach. In the latter part of this section, we also introduce tensor network diagrams~\cite{orus2019tensor}, which provide a graphical formalism for representing quantum circuits and are directly leveraged in our framework.

\subsection{Quantum Circuit Building Blocks}

\textbf{Qubits.} Qubits are the fundamental units for quantum information processing. Each qubit is represented by its state, which is defined as
\begin{align} 
|\psi\rangle = \alpha|0\rangle + \beta|1\rangle, 
\end{align}
where $\alpha$ and $\beta$ are complex numbers known as amplitudes with the property that $|\alpha|^2 + |\beta|^2 = 1$ and $|0\rangle$ and $|1\rangle$ represent an orthonormal basis set. Quantum states of multiple qubits can be composed using tensor products~\cite{nielsen2010quantum}. For example, for a two-qubit system, we can represent the state vector for that system as
\begin{align}
    |\psi_{AB}\rangle = |\psi_A\rangle \otimes |\psi_B\rangle, 
\end{align}
where $\otimes$ represents the Kronecker product of two vectors or matrices~\cite{horn1990matrix}. The orthonormal basis set for the two qubits is also expressed using the Kronecker product. For example, the basis set $|01\rangle$ for the two qubits is expressed as
\begin{align}
    |01\rangle = |0\rangle\otimes|1\rangle = \begin{bmatrix}1\\0\end{bmatrix}\otimes\begin{bmatrix}0\\1\end{bmatrix} = \begin{bmatrix}0\\1\\0\\0\end{bmatrix}.
\end{align}
In the above equation, we view the basis set as a one-dimensional column vector of size $4$. However, by stacking the one-dimensional vector using a column major layout, we can view the basis set as a two-dimensional matrix such that
\begin{align}
    |01\rangle = \begin{bmatrix}0 & 0 \\ 1 & 0\end{bmatrix}.
\end{align}
Similarly, the state vector for the two-qubit system can be viewed as a one-dimensional column vector of size four, or it can be viewed as a two-dimensional tensor with each dimension being equal to two. In general, for a {\it{$d$ qubit system}} the state vector and the corresponding basis sets can be viewed as a one-dimensional column vector of size $2^d$, or it can be viewed as a $d$-dimensional tensor where each dimension is of size two. Figure~\ref{fig:basics} a) outlines three tensors of different dimensions with their corresponding tensor diagram notation, also called Penrose graphical notation~\cite{orus2014practical}. A tensor diagram represents the tensors as nodes with edges showing the dimension of the tensors. We will use this notation to represent quantum circuits in later sections.%

\begin{figure}[t]
    \centering  
    \includegraphics[width=0.5\linewidth]{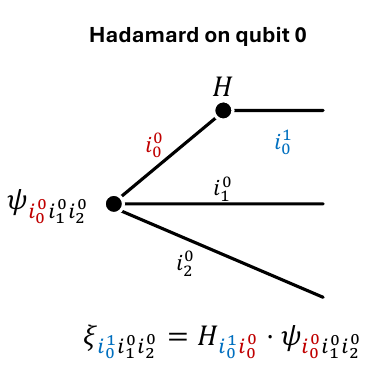}
    \caption{The graph representation of the tensor contraction between a two dimensional tensor $H$ and a three dimensional tensor $\psi$. The result is the three dimensional tensor $\xi$. The nodes represent the tensors, while the edges represent the dimensions. The dimension with the red index specifies the contraction dimension.}
    \label{fig:tensor_contractions}
    \vspace{-3mm}
\end{figure}

{\bf{Quantum Gates.}} Quantum gates serve as the fundamental building blocks of quantum circuits, implementing the unitary operations that manipulate qubit states. These operations are mathematically described as unitary matrices relative to some orthonormal basis set.
Figure~\ref{fig:basics} c) provides examples of single- and two-qubit quantum gates used in quantum algorithms like Quantum Fourier Transform~\cite{shor1997polynomial}, Grover's algorithm~\cite{grover1996fast}, Variational Quantum Eigensolvers~\cite{peruzzo2014variational}, etc.
The simplest gates operate on individual qubits and are represented as $2 \times 2$ unitary matrices. More complex operations require multiple qubits, such as controlled gates that apply an operation conditionally based on the state of a control qubit, or swap gates that exchange the states of two qubits. While these multi-qubit gates can still be expressed as matrices—for instance, a two-qubit controlled gate becomes a $4 \times 4$ matrix.
—they can also be conceptualized as higher-dimensional tensors. A controlled gate operating on two qubits naturally corresponds to a four-dimensional tensor. More generally, any quantum gate acting on $p$ qubits can be represented as a $2p$-dimensional tensor, with each dimension of size two.

{\bf{Quantum Circuits.}} Quantum circuits are composed of quantum gates that are applied sequentially on the qubits, followed by a measurement step. During gate operations, qubits remain in superposition, but measurement causes the quantum states to collapse to either $0$ or $1$ with specific probabilities. Figure~\ref{fig:basics}b) shows a simple circuit with three qubits and three gates. The Hadamard gate is applied on qubit $0$. The first controlled X gate is applied on qubit $1$ but it is conditioned on qubit $0$. The second controlled gate is applied on qubit $2$ and controlled by qubit $1$.

While individual gates target specific qubits, their effect on the quantum state vector requires $2^d\times 2^d$ matrix operations. This is achieved through tensor products: when a gate acts on one qubit while others remain unchanged, we Kronecker product the gate with identity matrices. For example, the Hadamard gate applied on qubit $0$ can be expressed as an $2^3\times 2^3$ matrix such as
\begin{align}
    G_H = H\otimes I_2\otimes I_2, 
\end{align}
and the first control X gate applied on qubits $0$ and $1$ can also be represented as a tensor product such that
\begin{align}
    G^0_{CX}=CX\otimes I_2,
\end{align}
where $CX$ is represented as a $4\times 4$ matrix. 

Entire circuits can be represented using this matrix-vector notation. However, implementing the computation is never done as dense matrix multiplications. The implementation must explore the structure of these matrices to reduce the amount of computation, since all Kronecker product matrices are sparse. Current work such as LFSQ~\cite{tarabkhah2025synthesis} exploits these properties when optimizing quantum circuits and provide a domain specific language to generate quantum circuits via compilation. The work provides a similar language to the SPL~\cite{xiong2001spl} notation, which has been widely utilized by the SPIRAL framework~\cite{franchetti2018spiral, mionis2020quantum} to generate linear transforms and more specifically Fourier transforms~\cite{popovici2018large, popovici2015generating}. In this work, we adopt a different direction, where we focus on developing a {\it{tensor-based framework}}. As outlined above, the state vector and the gates are tensors that are contracted in a specific order. The order will be decided based on our approach.

\begin{figure*}[t]
    \centering
    \includegraphics[width=0.8\textwidth]{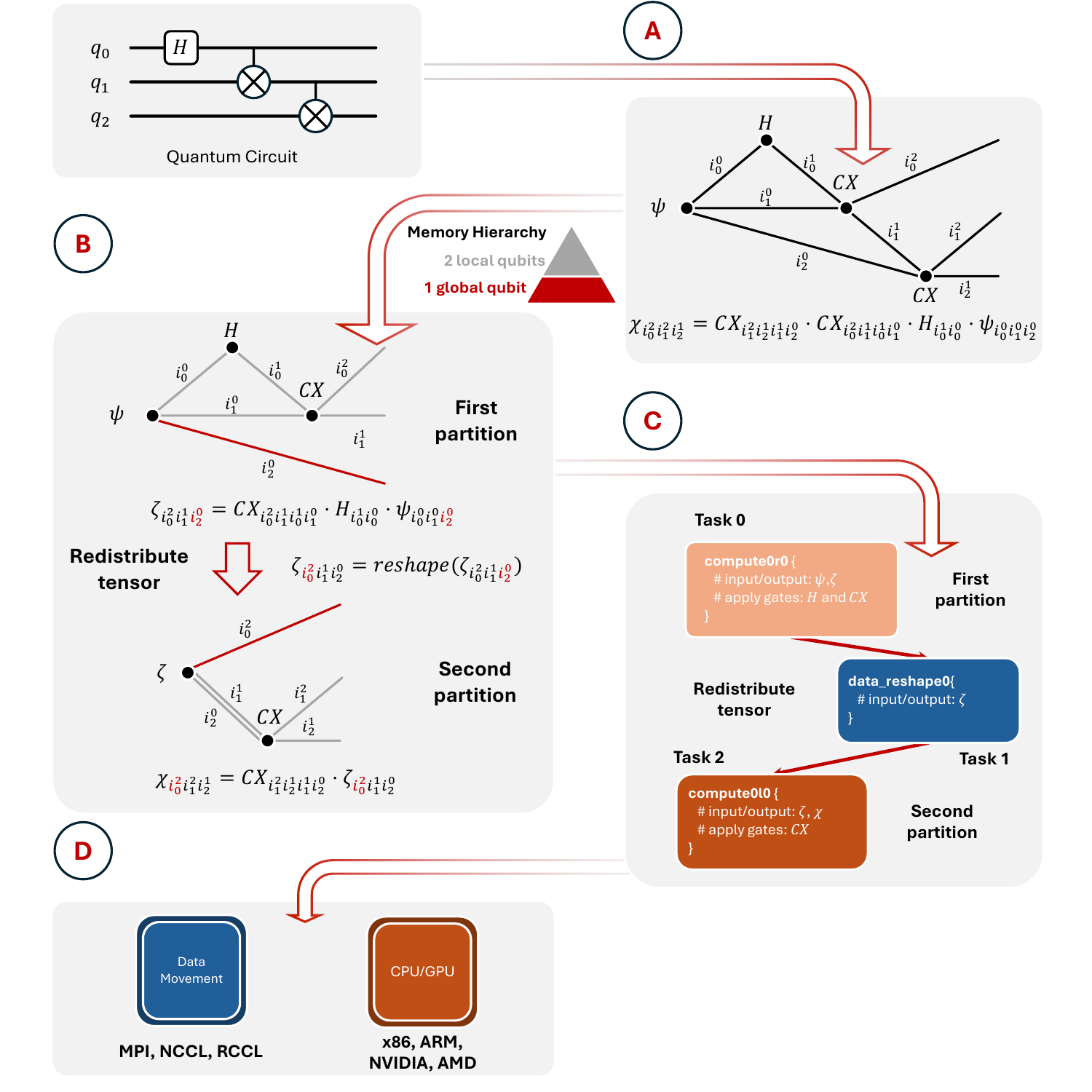}
    \caption{The overview of framework. The input is a quantum circuit, while the output is a C++ code that is compiled and can be executed on a specified platform. It translates the quantum circuit into a tensor contraction network (Step A). The network is traversed and for each node/tensor we compute the closeness centrality. Based on the specified memory hierarchy and its levels, the network is partitioned (Step B). For each partition the framework creates the computation and infers the data reshapes (Step C). The code is then lowered to the specified hardware platform (Step D), ready for compilation and execution.}
    \label{fig:overview}
    \vspace{-3mm}
\end{figure*}

\subsection{Tensors and Tensor Operations}

As discussed in the previous section, the state vector and quantum gates are multi-dimensional tensors. Hence quantum circuits can be viewed as chains of tensor contractions. 

Given $d$ qubits, the state vector is a $d$-dimensional tensor represented as
\begin{align}
    \psi_{i^0_0i^0_1\ldots i^0_{d-1}},
\end{align}
where each dimension $l=0..d-1$ is delineated by the index $i^0_{l}\in\{0, 1\}$. Note that the indices have both a subscript and a superscript. The subscript is meant to represent the corresponding dimension, i.e. dimension $l=0..d-1$. The superscripts are used to track the "time steps" or the stages of the computation when we apply the gates, which we explain next using our example in Figure \ref{fig:basics}.

Quantum gates applied on certain qubits are also tensors. Therefore, quantum circuits are chains of tensor contractions between the high dimensional state vector and the tensors representing the different gates. In our driving example, the Hadamard gate is applied on the qubit $0$. First, we represent the Hadamard gate as a two dimensional tensor
\begin{align}
    H_{i^1_0i^0_0},
\end{align}
and then the application of the Hadamard gate onto the state vector as a tensor contraction between a 3D tensor and a 2D tensor such that
\begin{align}   
\xi_{{\textcolor{blue}i}^{\textcolor{blue}1}_{\textcolor{blue}0}i^0_1i^0_2} = H_{{\textcolor{blue}i}^{\textcolor{blue}1}_{\textcolor{blue}0}{\textcolor{red}i}^{\textcolor{red}0}_{\textcolor{red}0}} \cdot \psi_{{\textcolor{red}i}^{\textcolor{red}0}_{\textcolor{red}0}i^0_1i^0_2}.
\end{align}
The $i^0_0$ index corresponds to qubit $0$, which is the \textit{contraction dimension} common between $\psi$ and $H$. Following Einstein summation convention, the repeated index $i^0_0$ appearing in both tensors is implicitly summed over during the contraction. The output tensor $\xi_{i^1_0i^0_1i^0_2}$ has index $i^1_0$ representing still qubit $0$, however with an altered state after applying the Hadamard gate. As such the superscripts keep track of the operations applied to the same qubits. 

A tensor contraction between two tensors can be represented as a contraction graph as depicted in Figure~\ref{fig:tensor_contractions}. As any graph, the contraction graph $\mathcal{G}$ is defined by the set of vertices or nodes $V$ and the set of edges $E$. For our case, the vertices $v_i\in V$ are tensors, e.g. the tensor representing the state vector and the tensors representing the quantum gates. The edge set $E$ represents the contraction dimensions between the tensors and the free dimensions that specify the output tensor dimensions. We will use this representation to capture quantum circuits. Moreover, we will develop our partitioning algorithm and subsequently the code generator on this notation and representation.

\section{From Quantum Circuit to Efficient Code}
\label{sec:method}

In this section, we present our approach of translating quantum circuits to contraction graphs, partitioning the operations and finally generating code. We briefly present tensor creation. We then dive into the partitioning algorithm. Finally, we provide details on our approach to generating the tasks required for the computation and data movement. Figure~\ref{fig:overview} provides an overview of our approach.

\subsection{From Quantum Circuit to Contraction Graph}

Any quantum circuit defines the number of qubits, regular registers and a list of quantum gates applied on the specific qubits. All quantum circuits end with a measurement stage, where the quantum state is collapsed to $0$ and $1$, with a specific probability. For example, the circuit definition of the driving example in Figure~\ref{fig:basics} is described as
\begin{center}
\begin{minipage}{0.3\textwidth}
\begin{minted}[linenos,framesep=2mm]{python}
qreg q[3];
creg c[3];

h q[0];
cx q[0],q[1];
cx q[1],q[2];

measure q -> c;
\end{minted}
\end{minipage}
\end{center}
where line $1$ specifies the number of qubits, and line $2$ specifies the number of classical registers. The classical registers are used to store the values produced by measuring the qubits. Lines $4$ -- $6$ specify the gate operations - Hadamard followed by the two control X gates. Finally, line 8 specifies all qubits are measured and the results are stored in the \texttt{c} registers. As outlined in the above example, the specification of the quantum circuit presents the gates as a flat list showing their sequential application to the state vector.

\begin{figure}[t]
    \centering
    \includegraphics[width=0.7\linewidth]{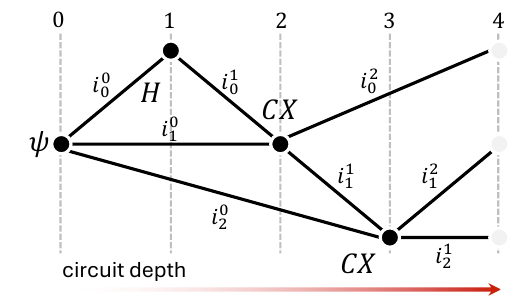}
    \caption{The contraction graph for the simple example with three qubits and three quantum gates. The nodes represent the tensors, the edges represent the dimensions (contraction dimensions and free dimensions). Each node is augmented with length of the critical path distance to the state vector $\psi$.}
    \label{fig:graph_structure}
    \vspace{-2mm}
\end{figure}

The framework parses either the \texttt{qasm} file or uses Qiskit~\cite{qiskit2024} to parse the \texttt{qasm} file and then traverse the internal data structure provided by Qiskit. As outlined in Figure~\ref{fig:overview}, step
\tikz[baseline=(char.base)]{\node[shape=circle,draw,inner sep=2pt] (char) {A}; } 
translates the quantum circuit and quantum gates into the defined contraction graph. Furthermore, the framework augments the nodes of the graph with extra information, namely the length of the critical path from the tensor representing the gate and the state vector. For each node $v_i$ we denote the critical path length as $l_{v_i}$. For example, the Hadamard $H$ gate is at $l_H = 1$, while the first $CX$ gate has $l_{CX} = 2$. There is a direct edge between the $CX$ tensor and state vector $\psi$. However, the computation cannot be performed until the tensor contraction with the $H$ gate is computed first. {\it{Any two nodes that share an edge will have dependent lengths of the critical path, while two tensors or nodes that do not have any common dimensions, are not on the critical path and can have independent lengths}}. The latter does not modify the sequential ordering of the tensor contractions, it just specifies that the two tensors can be contracted with the state vector in any order. For this work, we do not perform any sort of permutations or replacement of gates, we just traverse the gates as they are expressed in the original circuit. We leave this  optimization aspect as future work.

\subsection{Manipulating the Contraction Graph}

Large quantum circuit simulations require the state vector be distributed across multiple compute nodes. This implies that {\textit{communication is required between the compute nodes to fully simulate the circuit.}} Furthermore, modern hardware platforms have deep local memory hierarchies. Therefore, {\textit{the chain of tensor contractions must be clustered to exploit the faster memories}}. This is gate fusion mentioned in a multitude of papers~\cite{elsman2025gate, jones2019quest, 2023arXiv230801999B}, which in turn is loop fusion of matrix operations~\cite{raje2024const, dias2022sparselnr, dias2024sparseauto}. In this work, we propose a graph partitioning approach that centers around the idea of node centrality. In graph theory and network science, centrality refers to the relative importance or influence of a node within a graph. In our case, centrality will identify key nodes or tensors that are on the critical path of the computation. Previous works~\cite{gray2021hyper} have used {\bf{betweeness centrality}} to determine the criticality of nodes. However, computing betweeness ($BC$) centrality, which requires calculating all shortest paths that pass through each node, can become prohibitively expensive for a large number of nodes/quantum gates.

In this work, we focus on {\bf{closeness centrality}}. Closeness centrality ($CC$) measures how easily a node can reach all other nodes in the graph. It is calculated as the reciprocal of the average of the shortest path distances \( d(v_i,v_j) \): 
\begin{align} 
CC(v_i) = \frac{|RN(v_i)|}{dist(v_i)},
\end{align} 
where $RN(v_i)$ represents the reachable nodes, $|RN(v_i)|$ the size of $RN(v_i)$ and $dist(v_i) = \sum_{v_j\in RN(v_i)} d(v_i, v_j)$. In all our calculations, we use the modified the Wasserman and Faust formulation~\cite{} that multiplies the ratio by $\frac{|RN(v_i)|}{N}$ to account for possible disconnected components (most quantum circuits graphs may contain disconnected components). The CC captures a node's accessibility and requires knowledge of all reachable nodes and their distances. We develop a fast algorithm to keep track of all the reachable nodes and compute the distances of their critical path to the state vector tensor based on their dependencies. We then use these metrics to partition the graph as outlined in step 
\tikz[baseline=(char.base)]{\node[shape=circle,draw,inner sep=2pt] (char) {B}; }
in Figure~\ref{fig:overview}.

We start our algorithm by computing $v_i$'s reachable nodes and the distance from $v_i$ to all of them. Let $v_j\in N_d(v_i)$, where $N_d(v_i)$ is defined as the set of downstream neighboring nodes (nodes that have a critical path length higher than that of $v_i$'s). We initialize $RN(v_i)$ such that
\begin{align}
    RN(v_i) = \{v_j\} \cup RN(v_j)
\end{align}
and the distance from $v_i$ to all the reachable nodes from $v_j$ as
\begin{align}
    dist(v_i) = (1 + |RN(v_j)|) * |l_{v_i} - l_{v_j}| + dist(v_j),
\end{align}
where $|l_{v_i} - l_{v_j}|$ is the distance based on the lenght of the critical path. The algorithm iterates through all the nodes $v_k\in N_d(v_i)$ different from  $v_j$ and updates $RN(v_i)$ and $dist(v_i)$, accordingly. The distance $dist(v_i)$ is updated such as
\begin{align}
    dist(v_i) += (1 + |RN(v_k)|) * |l_{v_i} - l_{v_k}| + dist(v_k).
\end{align}
We need to account for the double counts, because $RN(v_i)\cap RN(v_k)$ may not be the empty set $\emptyset$. If $|RN(v_i)\cap RN(v_k)| < threshold$, where the threshold is defined, we subtract the distances between $v_i$ and all the nodes in $RN(v_i)\cap RN(v_j)$. If $|RN(v_i)\cap RN(v_j)| > threshold$, then we add the distances between $v_i$ and the nodes in 
$(RN(v_i)\setminus RN(v_k))\cup (RN(v_k)\setminus RN(v_i))$ and then divide the result by two. We construct an implementation that avoids redundant computation.

\begin{figure}[t]
    \centering
    \includegraphics[width=0.8\linewidth]{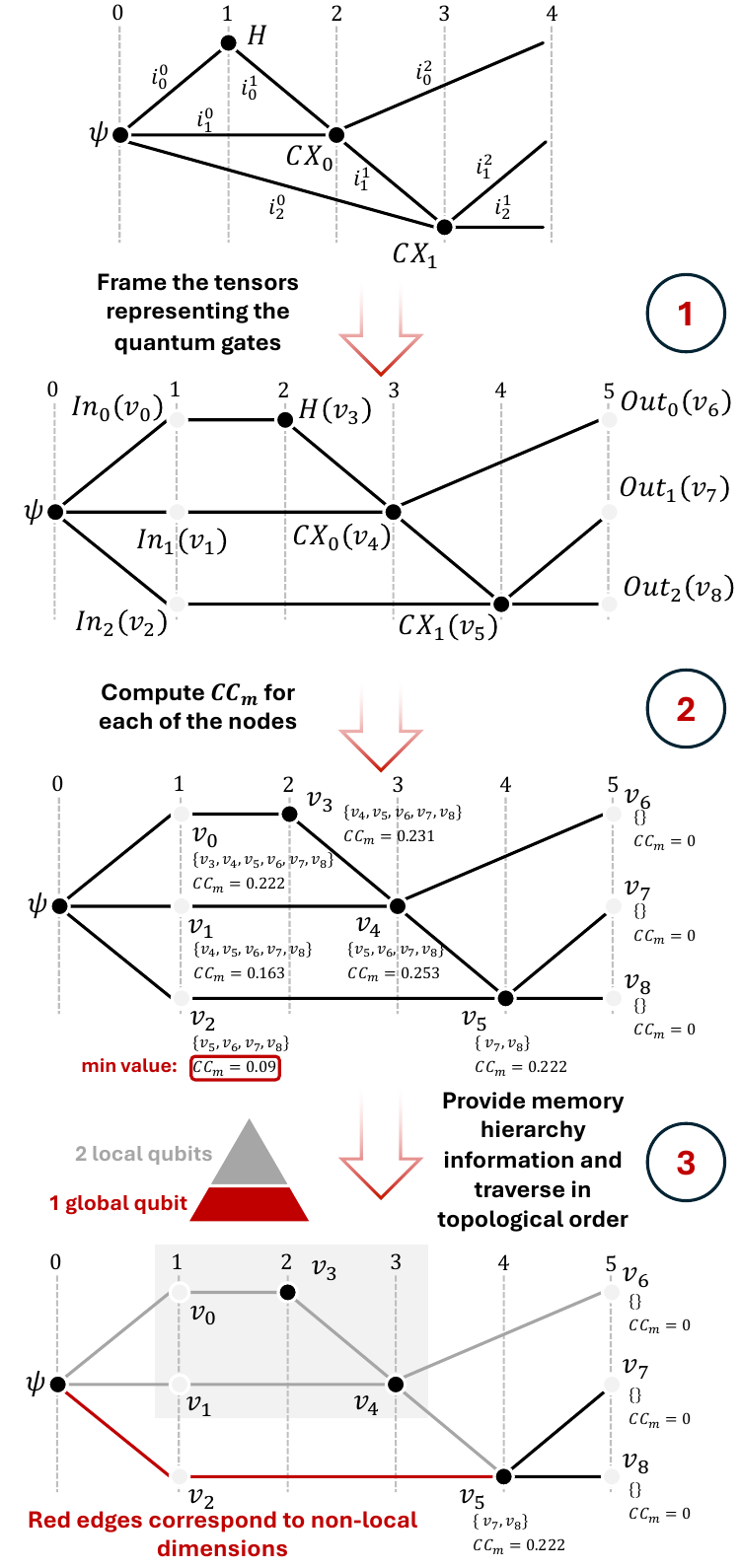}
    \caption{The steps required to compute the closeness centrality values for each of the tensor nodes, followed by the creation of the first partition based on information about the memory hierarchy (local and global qubits specify which dimensions of the state vector are kept local or not).}
    \label{fig:graph_manipulation}
    \vspace{-3mm}
\end{figure}

Figure~\ref{fig:graph_manipulation} outlines the steps taken by our algorithm to compute the closeness centrality for all the nodes and then to partition the contraction graph. 
Step
\tikz[baseline=(char.base)]{\node[shape=circle,draw,inner sep=2pt] (char) {1}; } creates some additional tensor nodes (gray nodes) around the original quantum gate tensors. These nodes are just $I_2$ identity matrices and will not modify the computation. The assumption is that these nodes are placed there as barrier points. Note that in Figure~\ref{fig:graph_manipulation}, we have relabeled some of the nodes for clarity. Step 
\tikz[baseline=(char.base)]{\node[shape=circle,draw,inner sep=2pt] (char) {2}; }
performs a backwards traversal of the graph, from the last nodes to the beginning. We gradually construct the list of reachable nodes $RN(v_i)$ and using our recursive definition for the centrality, augment each node with the corresponding values. Step 
\tikz[baseline=(char.base)]{\node[shape=circle,draw,inner sep=2pt] (char) {3}; }
performs a forward traversal of the graph. Additional information is provided, namely information about the memory hierarchy (the size of the local memories). The forward pass investigates the group of leftmost gray nodes and identifies the nodes with highest centrality that can be stored in the defined memory hierarchy. Then the algorithm moves this gray frontier until nodes or tensor contractions that depend on the dimensions that are not local are met. At that point, the algorithm creates a partition, readjusts the frontier nodes and repeats step 
\tikz[baseline=(char.base)]{\node[shape=circle,draw,inner sep=2pt] (char) {3}; } until the entire contraction graph is traversed and partitioned. Depending on the number of levels in the memory hierarchy the algorithm can be invoked for each of the newly determined partitions to further decompose the tensor operations and map the contractions to the given memory hierarchy.

\begin{figure*}[ht]
    \centering
    \includegraphics[width=0.9\textwidth]{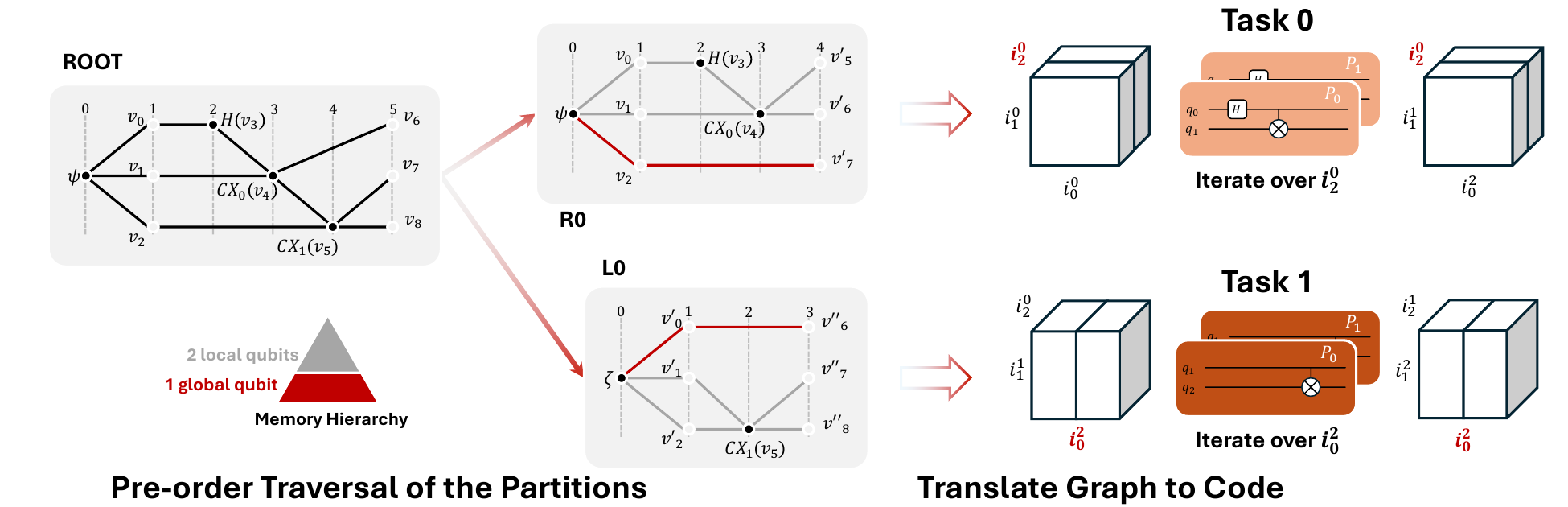}
    \caption{The partitions are stored as a tree, with two levels similar to the specified memory hierarchy. The tree is traverse using a pre-order traversal. Step C creates the corresponding compute and data movement tasks as outlined in Subsection~\ref{sub:graph_to_code}.}
    \label{fig:graph_to_code}
\end{figure*}

When performing the forward pass, we proceed with the traversal until we reach a node or a contraction that has a dimension that is not local (red edge). There are two cases where the forward will continue
\begin{itemize}
    \item The contraction tensor has only elements on the main diagonal different from $0$. For example, a phase gate applied one single qubit is defined as $P(\alpha) = \begin{bmatrix} 1 & 0\\0 & e^{-j\alpha}\end{bmatrix}$. The off-diagonal elements are $0$. 
    \item The contraction tensor is a controlled gate, and the dimension that corresponds to the control is the one that is not local. Therefore, the forward pass can proceed on the dimension corresponding to the control.
\end{itemize}
For all cases, the forward pass continues the grouping operation as long as the dependencies between the gates are satisfied. In other words, the algorithm will not break the correctness of applying the gates to the state vector compared to the sequential implementation.

\subsection{From Partitions to Code}
\label{sub:graph_to_code}

As outlined in Figure~\ref{fig:overview}, step
\tikz[baseline=(char.base)]{\node[shape=circle,draw,inner sep=2pt] (char) {C}; } parses the partitions and translates the representation into the {\it{buffers}} that store the state vector, the {\it{data movement}} that partitions and reshapes the state vectors to fit the defined memory hierarchy, and {\it{computation}} that applies the small quantum gate tensors on the state vector. We create a simple algorithm that traverses the partitions stored as a tree data structure. Figure~\ref{fig:graph_to_code}, shows the tree structure used to store the partitions obtained for our three qubit example in Figure~\ref{fig:basics}b). The partition tree has two levels, the same as the memory hierarchy ($2$ local qubits, and $1$ global qubit). The description of a deeper memory hierarchy would have created a taller partition tree. In our example, the tree has one root node and two children. Each node contains information about the contraction graph and the dimensions/edges that are kept local or global based on the defined memory hierarchy. Using this information and performing a pre-order traversal, we can translate the partition tree into tasks and code.

{\bf{Partitioning the Input/Output Tensors.}} Recall that each contraction graph specifies the operations between the state vector and the small quantum gate tensors. For example, the \texttt{ROOT} node in Figure~\ref{fig:graph_to_code} contains the entire contraction graph, while the children nodes \texttt{L0} and \texttt{R0} contain the partitioned contraction graphs based on the two level memory hierarchy. Based on the decorated dimensions as global dimensions, the state vectors will be partitioned accordingly. For example, in partition node \texttt{L0} the $i^0_2$ dimension is marked as global. Therefore, the corresponding state vectors $\psi_{i^0_0i^0_1i^0_2}$ and $\zeta_{i^2_0i^2_1i^0_2}$ will have the $i^0_2$ dimension partitioned. Similarly, in node \texttt{R0} the $i^2_0$ dimension is marked as global and $\zeta_{i^2_0i^2_1i^0_2}$ and $\chi_{i^2_0i^2_2i^1_2}$ will have the $i^2_0$ dimension partitioned. Partitioning the large tensors means that the computation is performed in blocks or tiles, and smaller buffers may be needed to store the partitions.

{\bf{Creating the Data Reshaping Operations.}} Data must be moved from large buffers into smaller buffers before applying the computation. As mentioned above, tensors that represent the quantum state vector are partitioned on specific dimensions. Depending on the memory hierarchy different types of operations are needed. If the tensors are partitioned and distributed across multiple compute nodes, then data must be reshaped between stages of the computation. For example, node \texttt{L0} requires the output tensor $\zeta_{i^2_0i^2_1i^0_2}$ be partitioned on $i^0_2$, while node \texttt{R0} requires the input tensor $\zeta_{i^2_0i^2_1i^0_2}$ partitioned across $i^2_0$. Given the mismatch in partitioned dimensions for the same buffer $\zeta$, a communication over the network is triggered. Our traversal algorithm identifies these cases and then generates the appropriate functions to pack/unpack the tensors and then communicate between compute units. Moreover, our partition algorithm attempts to identify the minimal number of such traversals. 

However, if the tensors are partitioned but kept on the memory of a single device, then packing and unpacking operations are used to load and store data to and from the memory hierarchies (caches on the CPU and shared memory on the GPU). In the case of single device, we define multiple levels of the memory. Astute readers will be familiar with the fact that most modern CPUs and GPUs have access to global memory (DRAM) and local memory (caches and shared memory). Under this scenario, the tensors are partitioned in such a way to create tiles or blocks that can be loaded from global memory and stored in local memory. The developed algorithm will analyze the partitioned dimensions, create the local buffers and then create the packing and unpacking functions to efficiently load the data from DRAM to local memory and use memory bandwidth most efficiently. For this scenario, we will use information about the local memories to better utilize the local memory hierarchy. We will provide more details in the experimental section.

{\bf{Creating the Computation.}} Computation always prefers data to be as close as possible to the cores. Partitioning the tensors, and then creating the data movement functions, allows for data to be reshaped and packed in fast local memory. Computation is generated based on the types of quantum gates. We base our work on the TACO compiler~\cite{kjolstad2017taco,chou2018format, ahrens2022autoscheduling} and its internal representation for tensor operations, namely the concrete index notation. We extend the work to represent the different small tensors applied on the large $d$-dimensional state vector. We construct primitive blocks that capture most of the computation. We automatically generate the necessary parts of the computation. Moreover, we perform classical loop optimizations such as loop fusion and loop unrolling to improve the execution time for the generated computation. Based on the computational platform we perform parallelization of the tensors. We exploit different levels of parallelism that best fits the underlying platforms.

Once the tasks are created and the code is specialized for the different hardware platforms, the code is compiled and executed on the preferred platform. We provide more details about the concrete implementations in the following section.

\section{Experimental Results and Discussion}
\label{sec:results}

In this section, we begin by describing the high-performance computing platforms used in our evaluation, including their architectural characteristics and relevance to our study. We then introduce the quantum circuit benchmarks selected to test our framework. We present performance results in comparison with baseline implementations. Finally, we discuss key optimization aspects, demonstrating the advantages of our approach while also identifying current limitations and directions for future work.

\subsection{Methodology}

\textbf{Evaluation Platforms.} Our experiments were conducted on three high-performance computing systems: Perlmutter, Frontier, Fugaku.

\begin{itemize}
    \item \textbf{Perlmutter (NERSC, USA):} Perlmutter is an HPE Cray Shasta system deployed at the NERSC. It consists of AMD EPYC “Milan” CPUs combined with NVIDIA A100 GPUs, interconnected by HPE Slingshot fabric. The system provides over 3,000 GPU nodes and delivers a peak performance of 70 petaflops (GPU partition). It is optimized for data-intensive and AI-driven scientific workloads~\cite{perlmutter2022nersc}.

    \item \textbf{Frontier (ORNL, USA):} Frontier, hosted at Oak Ridge National Laboratory, is the first exascale supercomputer, achieving over 1.1 exaflops on the HPL benchmark~\cite{top500_list}. It is built on HPE Cray EX architecture, with each node containing one AMD EPYC CPU and four AMD Instinct MI250X GPUs connected by Infinity Fabric. The system integrates HPE Slingshot interconnect and supports memory hierarchies with both high-bandwidth memory (HBM) and DDR4. Frontier represents the current state-of-the-art in heterogeneous, GPU-accelerated HPC systems~\cite{frontier2022ornl}.

     \item \textbf{Fugaku (RIKEN, Japan):} Fugaku, developed jointly by RIKEN and Fujitsu, is a CPU-only system based on ARM A64FX processors with 48 compute cores and high-bandwidth HBM2 memory. It achieved over 442 petaflops on the HPL benchmark, ranking first on the TOP500 list in 2020~\cite{top500_list}. Fugaku employs the Tofu-D interconnect and is optimized for memory-bound scientific workloads, offering a complementary architecture to GPU-accelerated platforms.
\end{itemize}

\textbf{Benchmarks.} To evaluate the performance and portability of the developed framework, we conducted experiments on a diverse set of quantum circuit benchmarks drawn from MQTBench\cite{wille2024mqtbench} and QASMBench\cite{li2021qasmbench}. These benchmark suites provide representative workloads that capture algorithmic patterns common in quantum computing and are widely used in performance evaluation studies. For our experiments, we executed benchmarks ranging from 30 to 37 qubits, which lie at the upper end of what is tractable for state-vector simulation on modern HPC systems. 

\begin{figure*}[t]
\centering

\begin{minipage}[b]{0.23\textwidth}
    \centering
    \includegraphics[width=\textwidth]{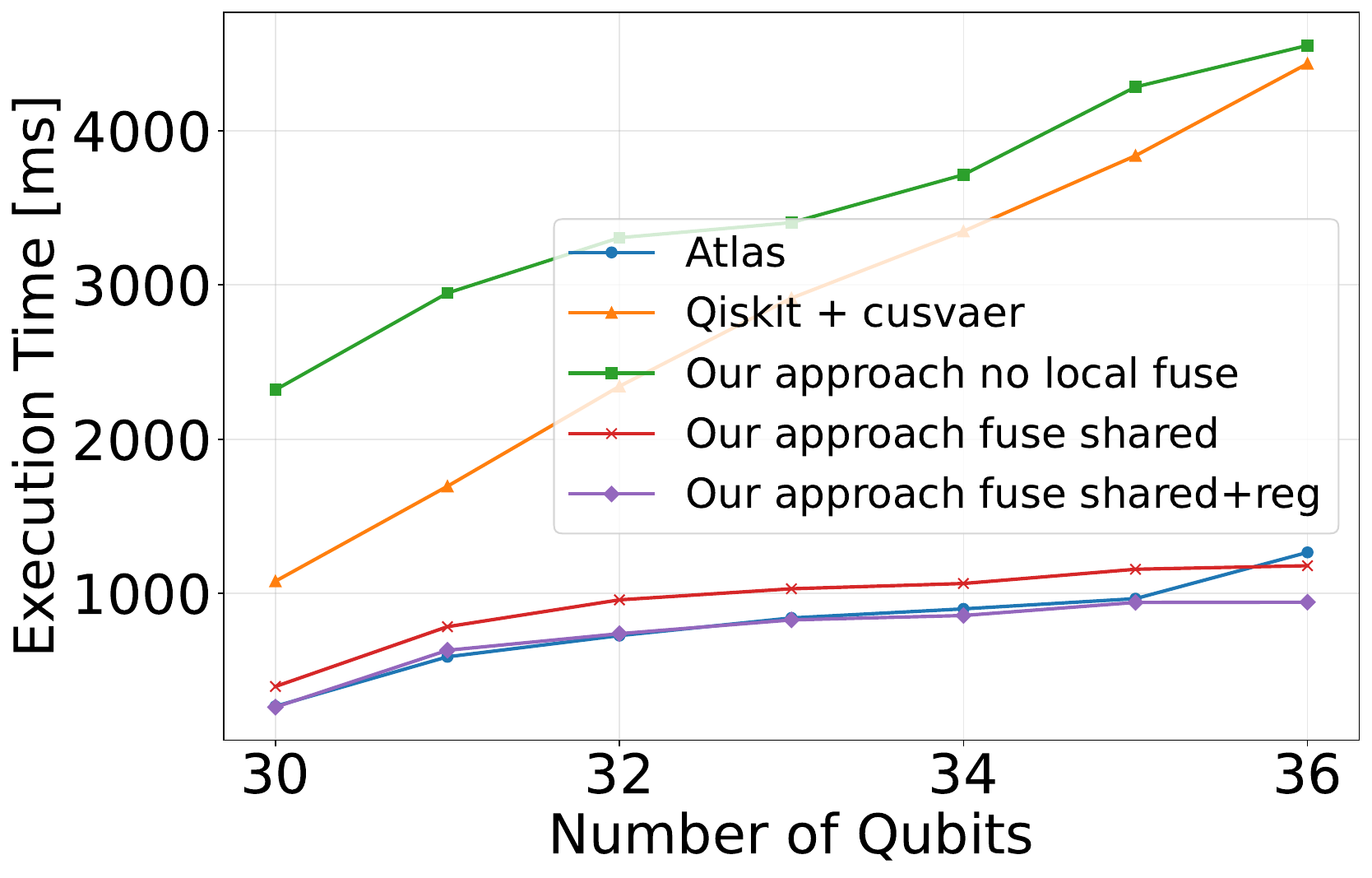}
    \caption*{AE Algorithm}
\end{minipage}
\hfill
\begin{minipage}[b]{0.23\textwidth}
    \centering
  '  \includegraphics[width=\textwidth]{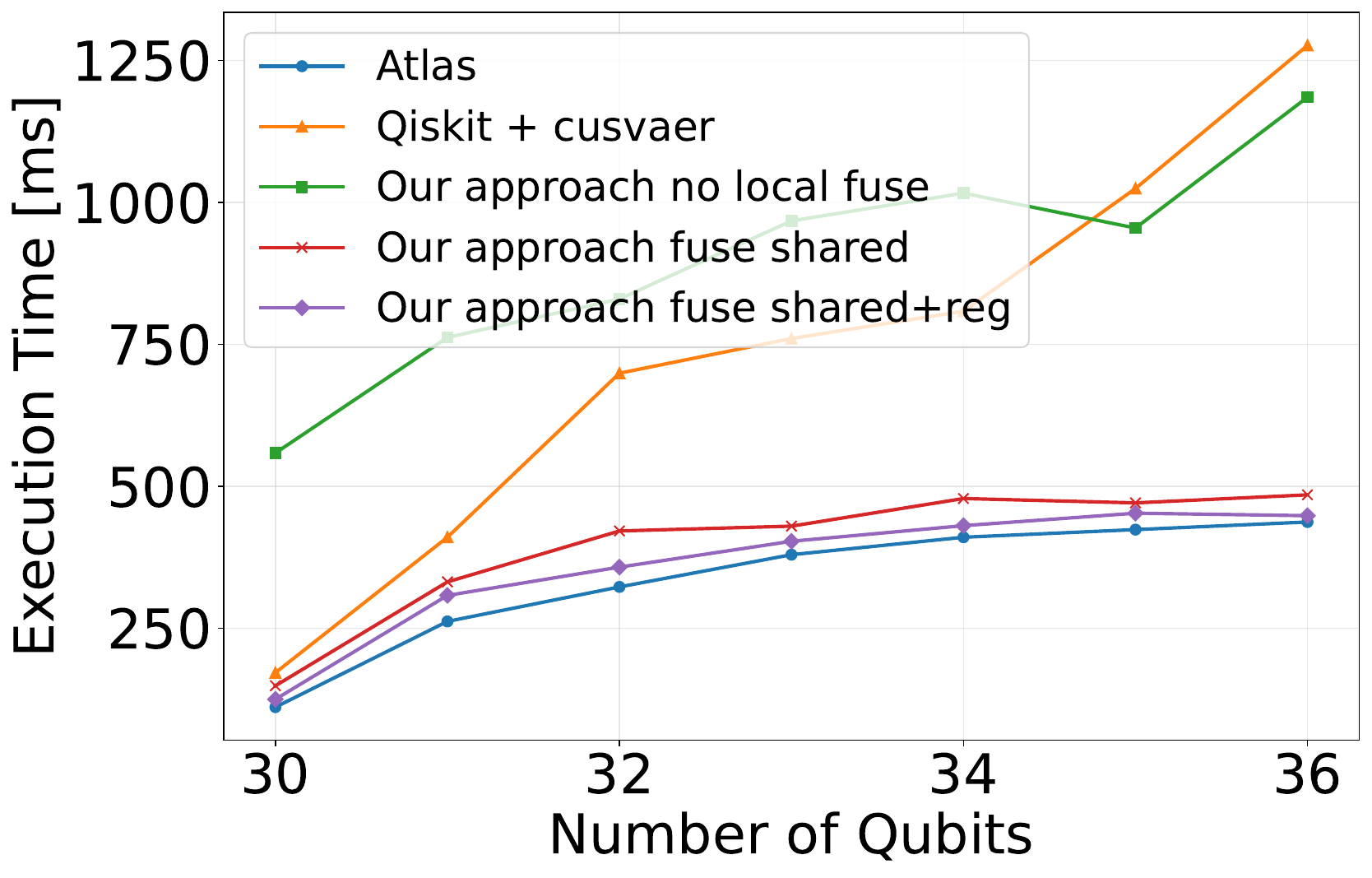}
    \caption*{DJ Algorithm}
\end{minipage}
\hfill
\begin{minipage}[b]{0.23\textwidth}
    \centering
    \includegraphics[width=\textwidth]{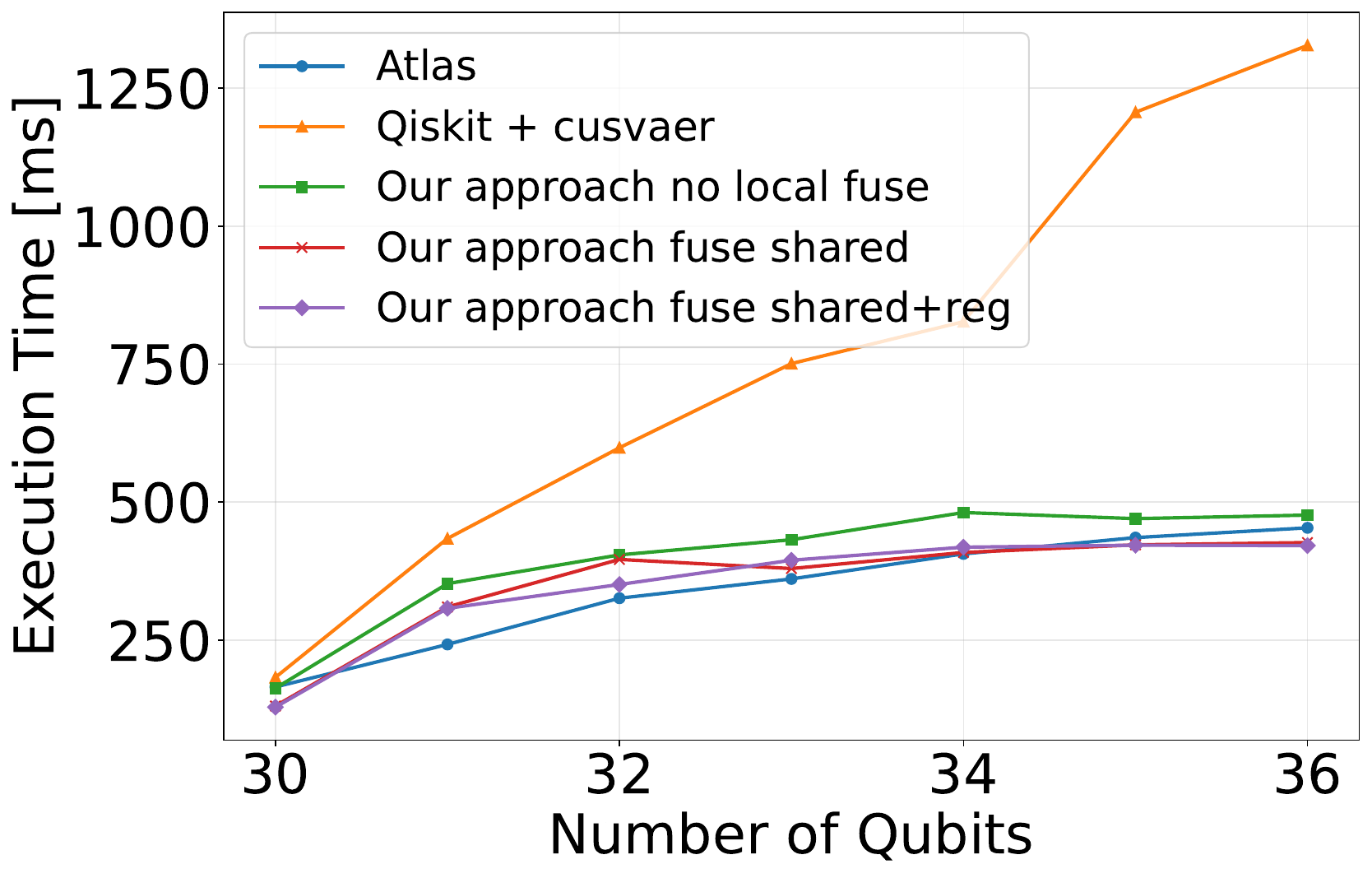}
    \caption*{GHZ Algorithm}
\end{minipage}
\hfill
\begin{minipage}[b]{0.23\textwidth}
    \centering
    \includegraphics[width=\textwidth]{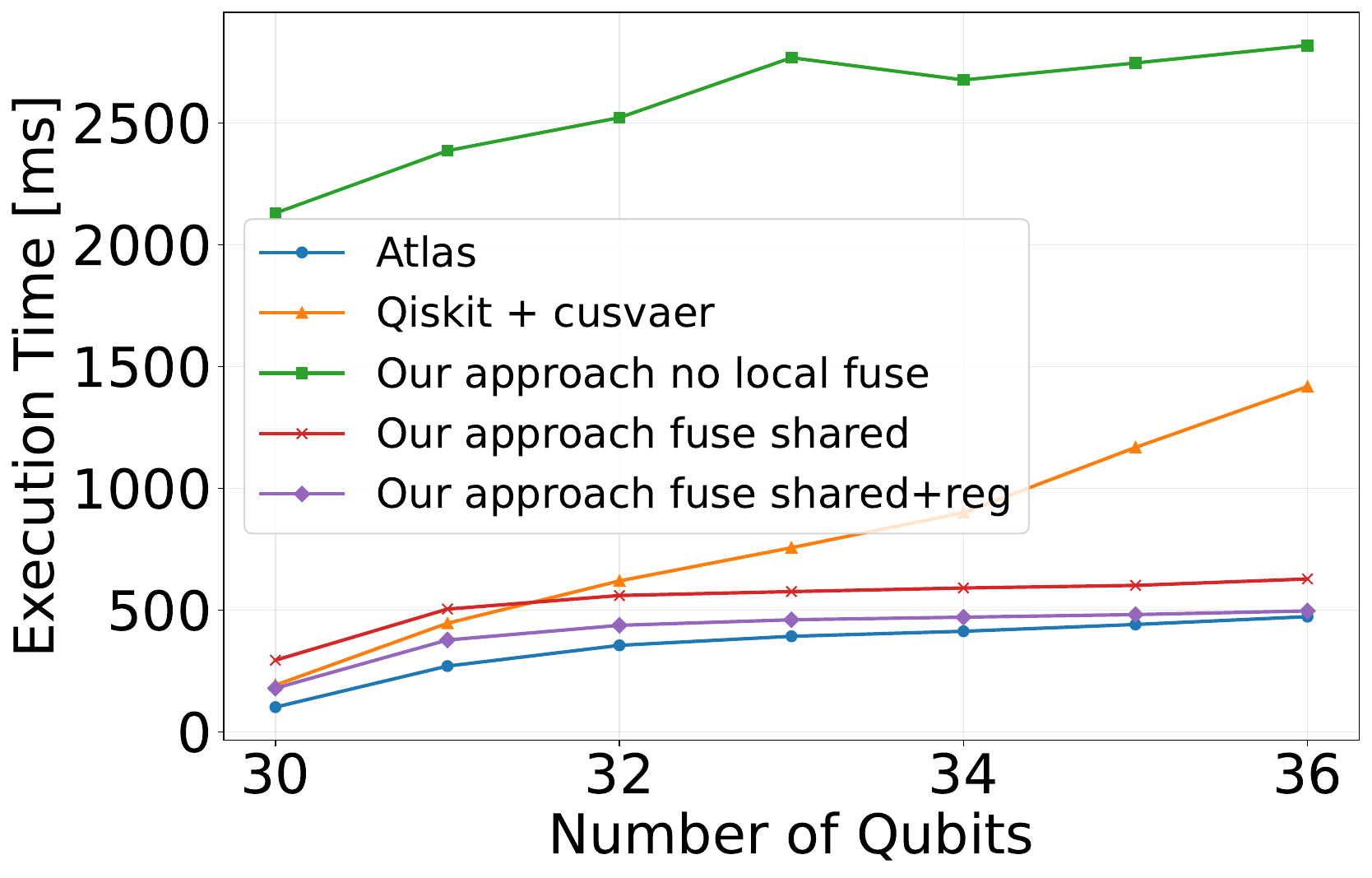}
    \caption*{Ising Algorithm}
\end{minipage}\\[0.5cm]

\begin{minipage}[b]{0.23\textwidth}
    \centering
    \includegraphics[width=\textwidth]{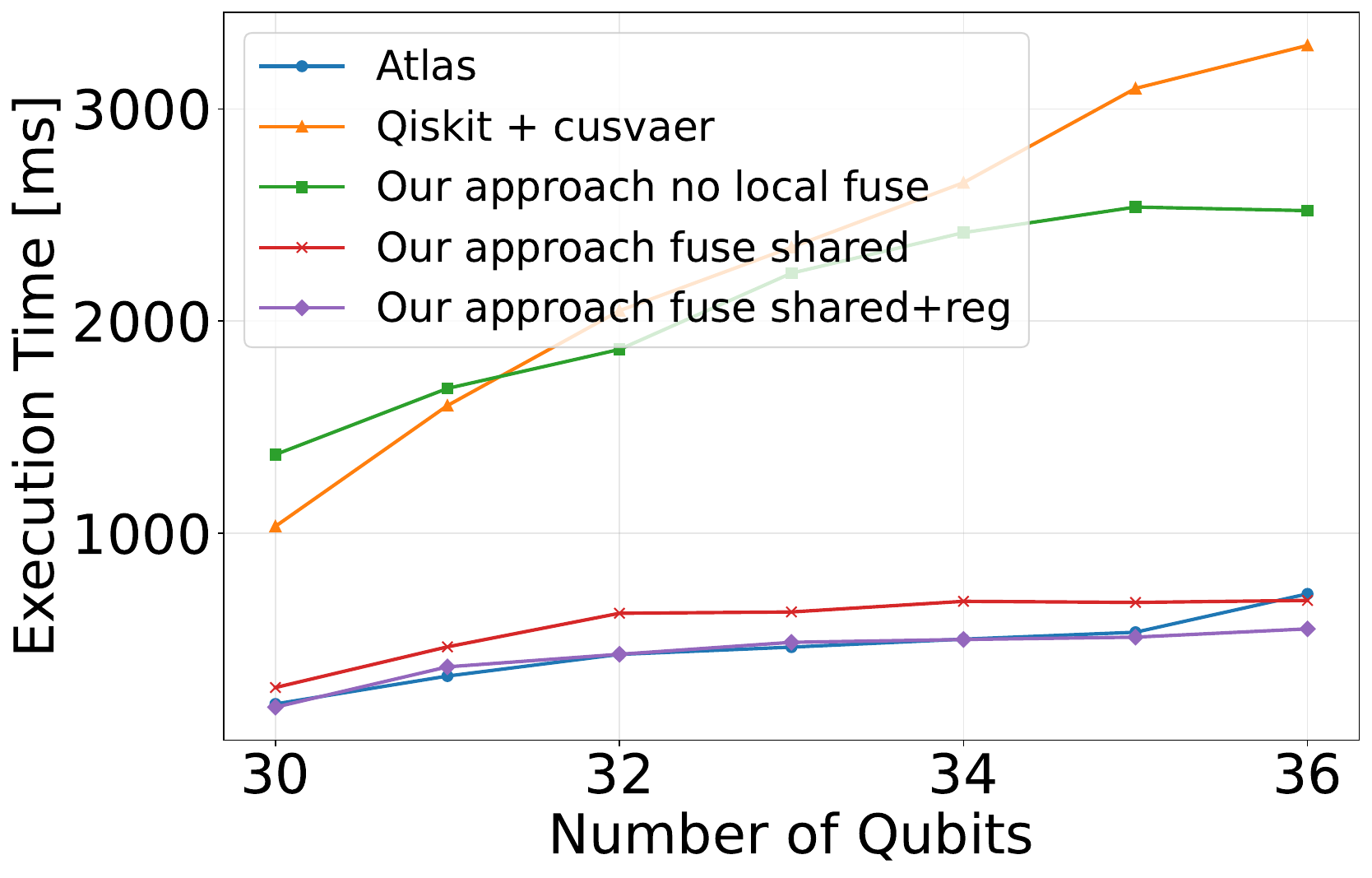}
    \caption*{QFT Algorithm}
\end{minipage}
\hfill
\begin{minipage}[b]{0.23\textwidth}
    \centering
    \includegraphics[width=\textwidth]{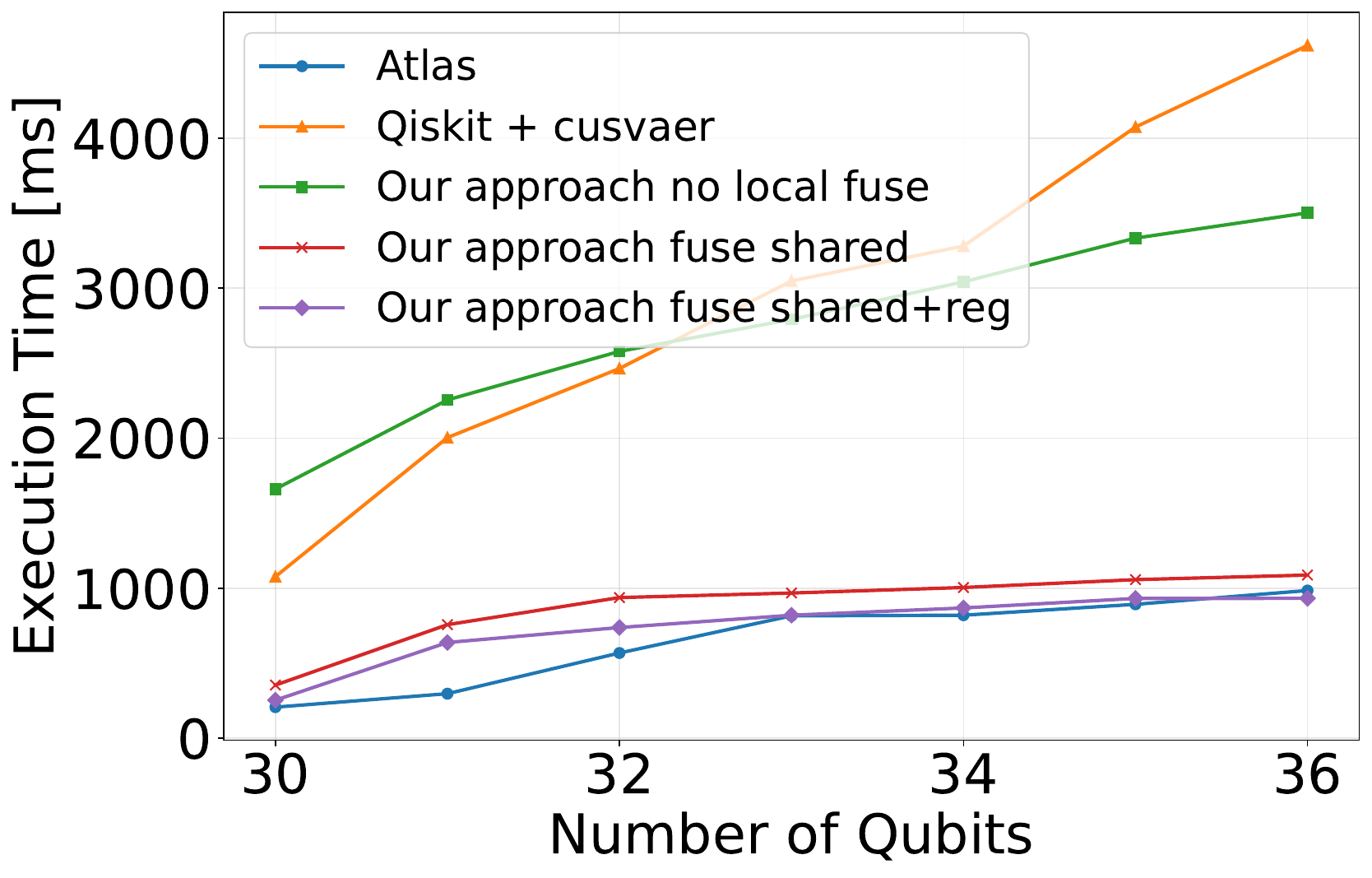}
    \caption*{QPE Exact Algorithm}
\end{minipage}
\hfill
\begin{minipage}[b]{0.23\textwidth}
    \centering
    \includegraphics[width=\textwidth]{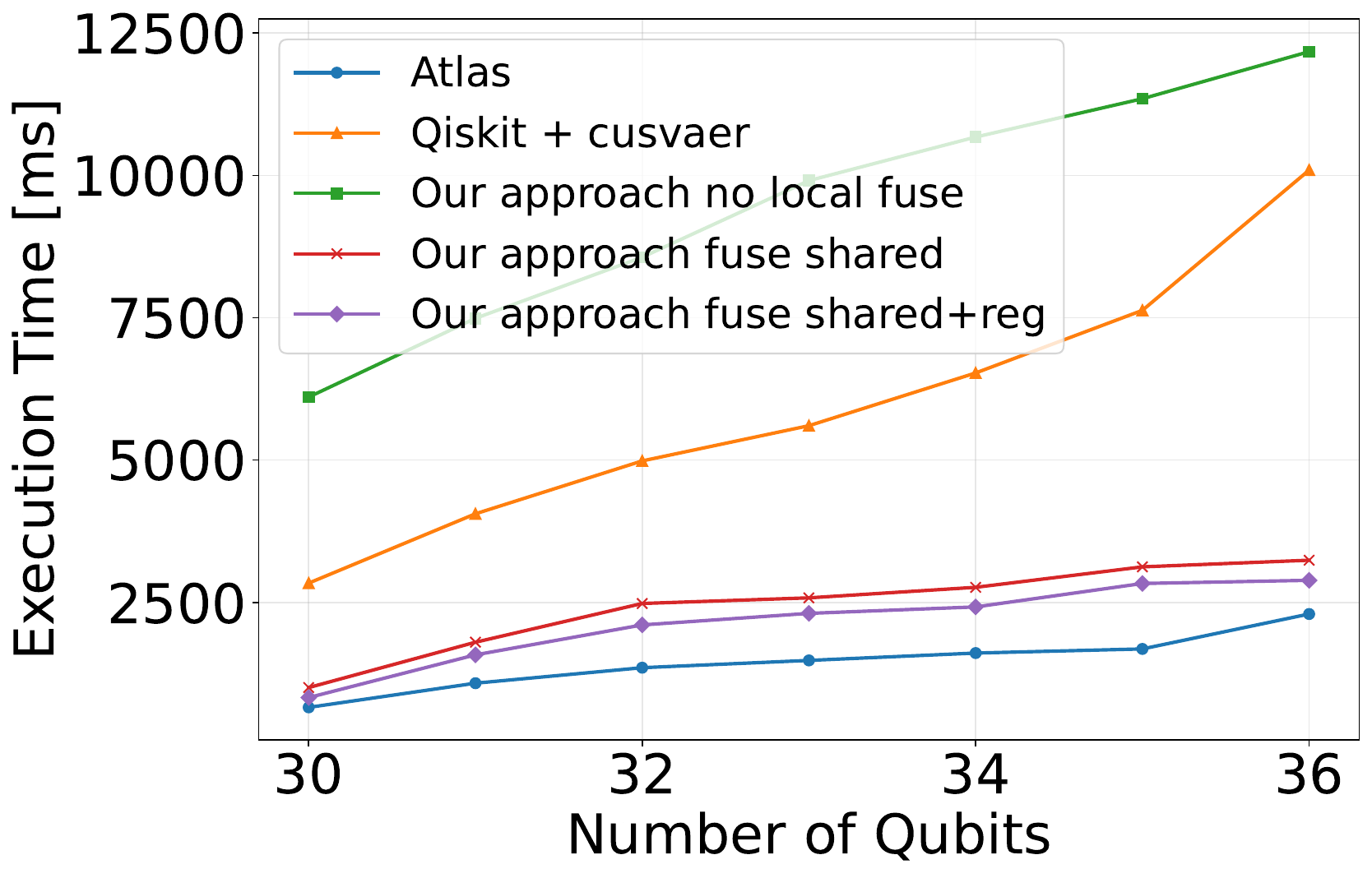}
    \caption*{SU2 Random Algorithm}
\end{minipage}
\hfill
\begin{minipage}[b]{0.23\textwidth}
    \centering
    \includegraphics[width=\textwidth]{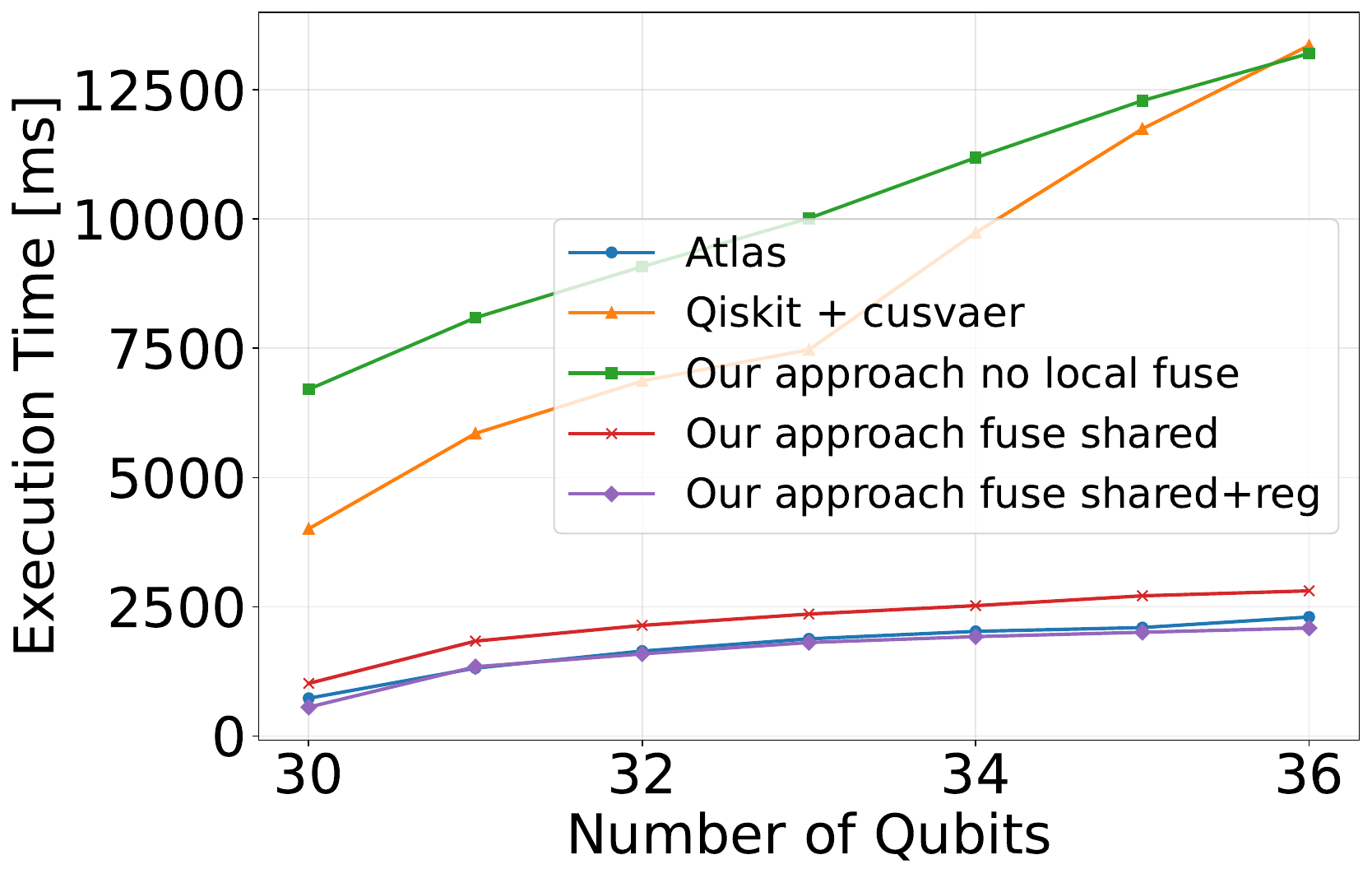}
    \caption*{VQC Algorithm}
\end{minipage}

\caption{Weak scaling results for eight quantum algorithms using Atlas~\cite{10.1109/SC41406.2024.00087}, cuQuantum~\cite{2023arXiv230801999B}, and our approach using no local gate fusion, gate fusion on shared memory, and finally gate fusion on both shared memory and registers.}
\label{fig:all_algorithms}
\end{figure*}

\begin{itemize}
    \item \textbf{MQTBench:} MQTBench provides a broad collection of circuits designed to stress-test quantum circuit simulators, including structured algorithms (e.g., Quantum Fourier Transform and Grover’s search), variational circuits, and random circuit families.

    \item \textbf{QASMBench:} From QASMBench, we selected the Ising model and Variational Quantum Circuit (VQC) benchmarks, which capture key application domains: physics-inspired Hamiltonian simulation and variational hybrid algorithms for near-term quantum computing.
\end{itemize}

\begin{figure}[t]
    \centering
    \includegraphics[width=0.65\linewidth]{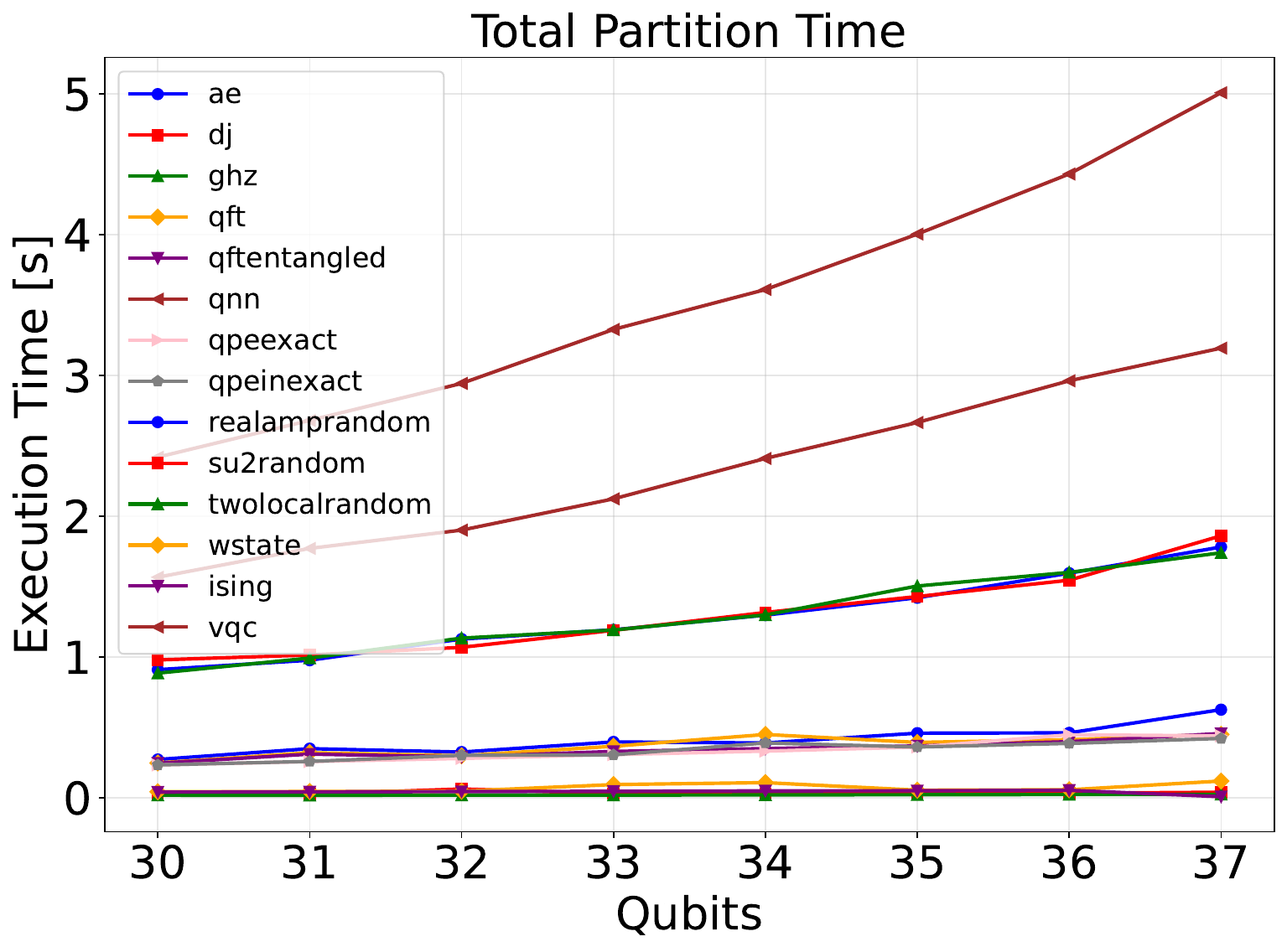}
    \caption{The overall execution time of our partitioning algorithm for the space of all quantum algorithms and number of qubits. For the \texttt{qnn} circuit our algorithm takes a maximum of $5$ seconds for $37$ qubits.}
    \label{fig:partition_time}
    \vspace{-3mm}
\end{figure}

By combining benchmarks from MQTBench and QASMBench, our evaluation covers a wide spectrum of circuit characteristics ranging from highly structured to randomized, and physics-inspired to hybrid variational. This diversity ensures that the performance of the proposed framework is assessed under realistic and heterogeneous workloads.

\textbf{Software and Baseline Implementations.} We have built the framework as a Python-based tool. We chose Python for ease of use. The implementation uses the Qiskit framework and the NetworkX framework~\cite{hagberg2008exploring}. NetworkX is used to represent and partition the contraction graphs. As for the code generator, we implement from scratch the intermediate representation and the rules to transform the code. Finally, the generated code is C++, HIP~\cite{bauman2019introduction} and CUDA~\cite{garland2008parallel} code, depending on the platform. The generated code uses MPI for communication over the network. For the GPU experiments, we utilize the \texttt{cray-mpich/8.1.30}'s GPU-aware MPI collectives for both Perlmutter and Frontier. For the CPU, we use OpenMP to parallelize the loops. For the CUDA code we use \texttt{cudatoolkit/12.4}. For the AMD platform, we use \texttt{rocm/6.4.1}. For the Fugaku system, we use the proprietary compilation tools and communication libraries.

For the baseline implementations, we use Qiskit~\cite{qiskit2024} and cuQuantum~\cite{2023arXiv230801999B} via their \texttt{cusvaer} implementation. Moreover, we experiment with the Atlas framework~\cite{10.1109/SC41406.2024.00087} and also its available artifact~\cite{xu2024artifact}. We follow the instructions and installion steps as described in their documentation. We only use these baselines for the NVIDIA platform. For the AMD platforms, to the best of our knowledge, there are no quantum simulation frameworks. We did not test our implementation against Qiskit, because the Atlas framework has already outlined that the simulations take a long period of time.

\subsection{Experimental Results}

In the following, paragraphs we will present results obtained on the different systems, outlining the competitive results and the short time to partition the circuits. We will continue with results on the AMD platform. Finally, we will provide a breakdown of the execution time for the QFT use case. We ran our approach with all quantum circuits on all platforms. However, for the following paragraphs, we will present the most interesting aspects.

\begin{figure*}[htbp]
\centering

\begin{minipage}[b]{0.23\textwidth}
    \centering
    \includegraphics[width=\textwidth]{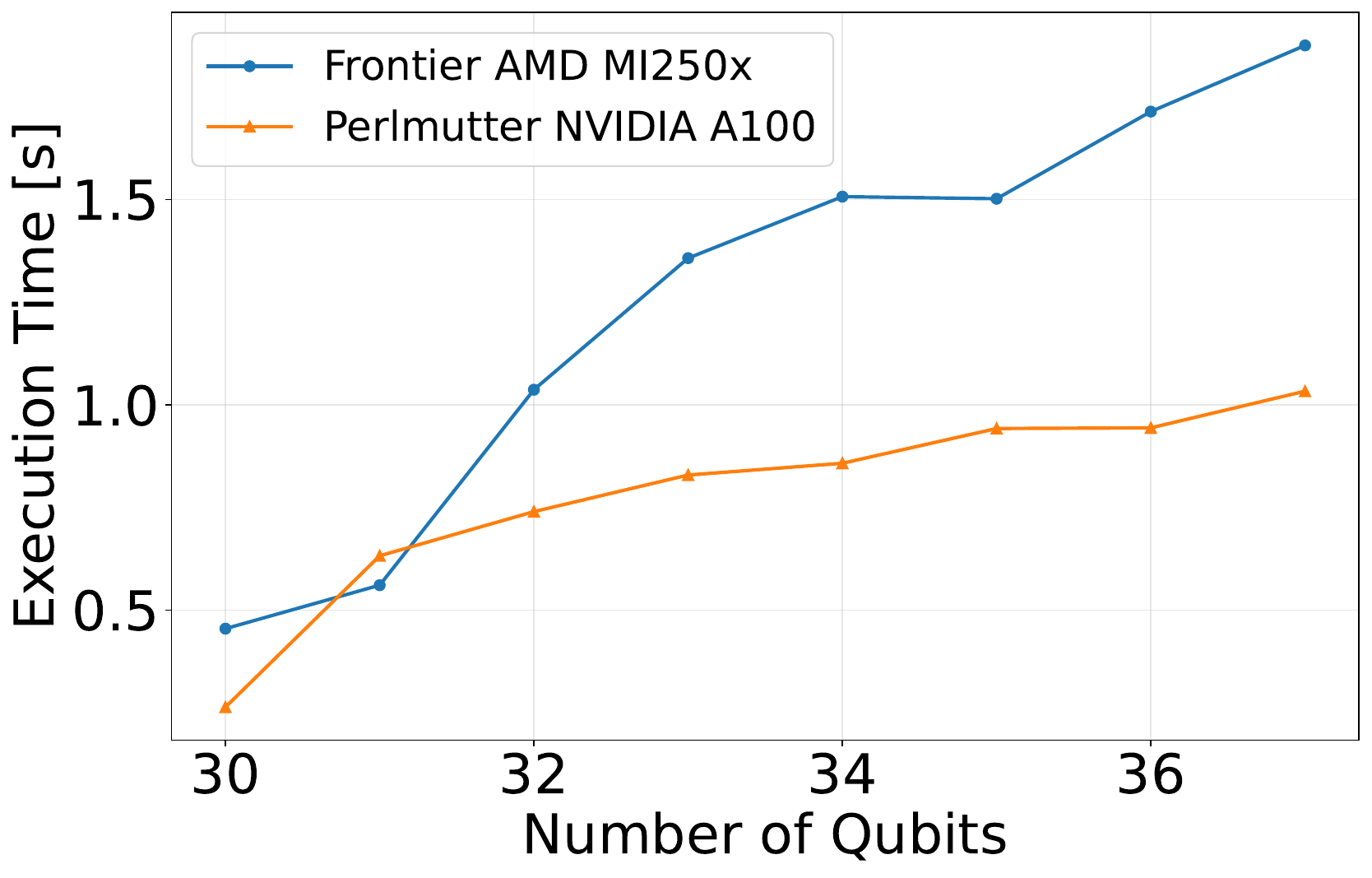}
    \caption*{AE Algorithm}
\end{minipage}
\hfill
\begin{minipage}[b]{0.23\textwidth}
    \centering
    \includegraphics[width=\textwidth]{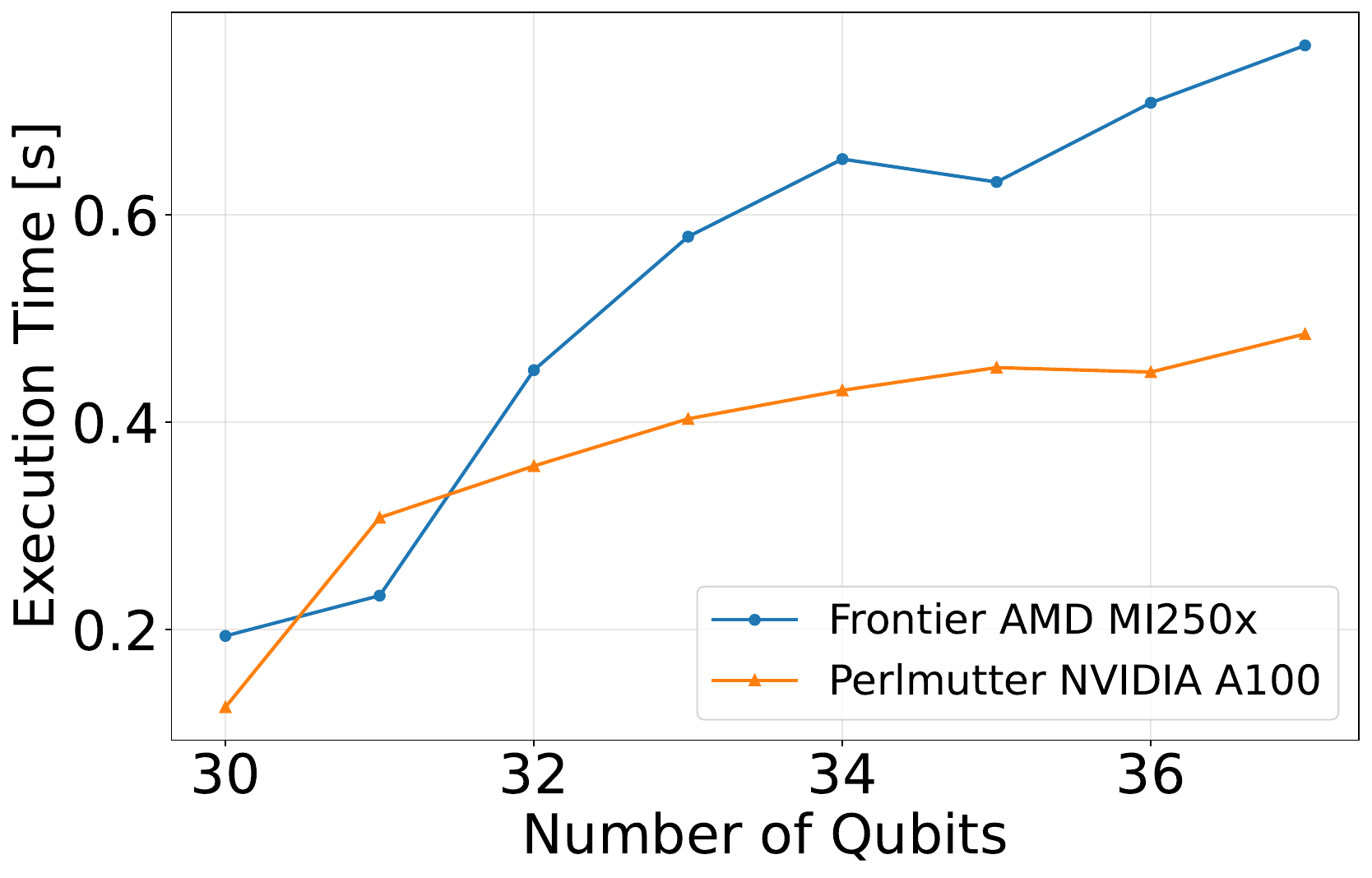}
    \caption*{DJ Algorithm}
\end{minipage}
\hfill
\begin{minipage}[b]{0.23\textwidth}
    \centering
    \includegraphics[width=\textwidth]{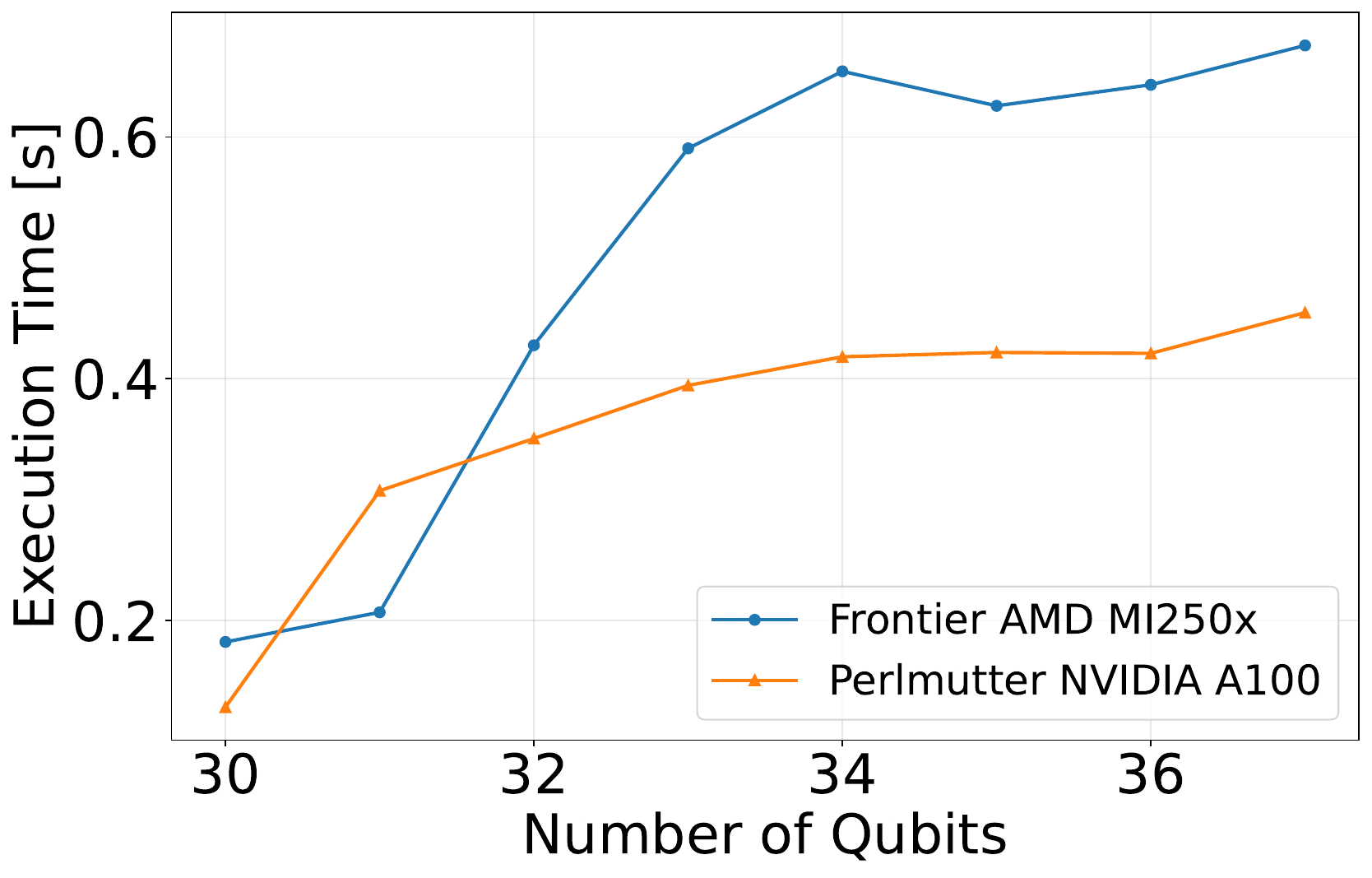}
    \caption*{GHZ Algorithm}
\end{minipage}
\hfill
\begin{minipage}[b]{0.23\textwidth}
    \centering
    \includegraphics[width=\textwidth]{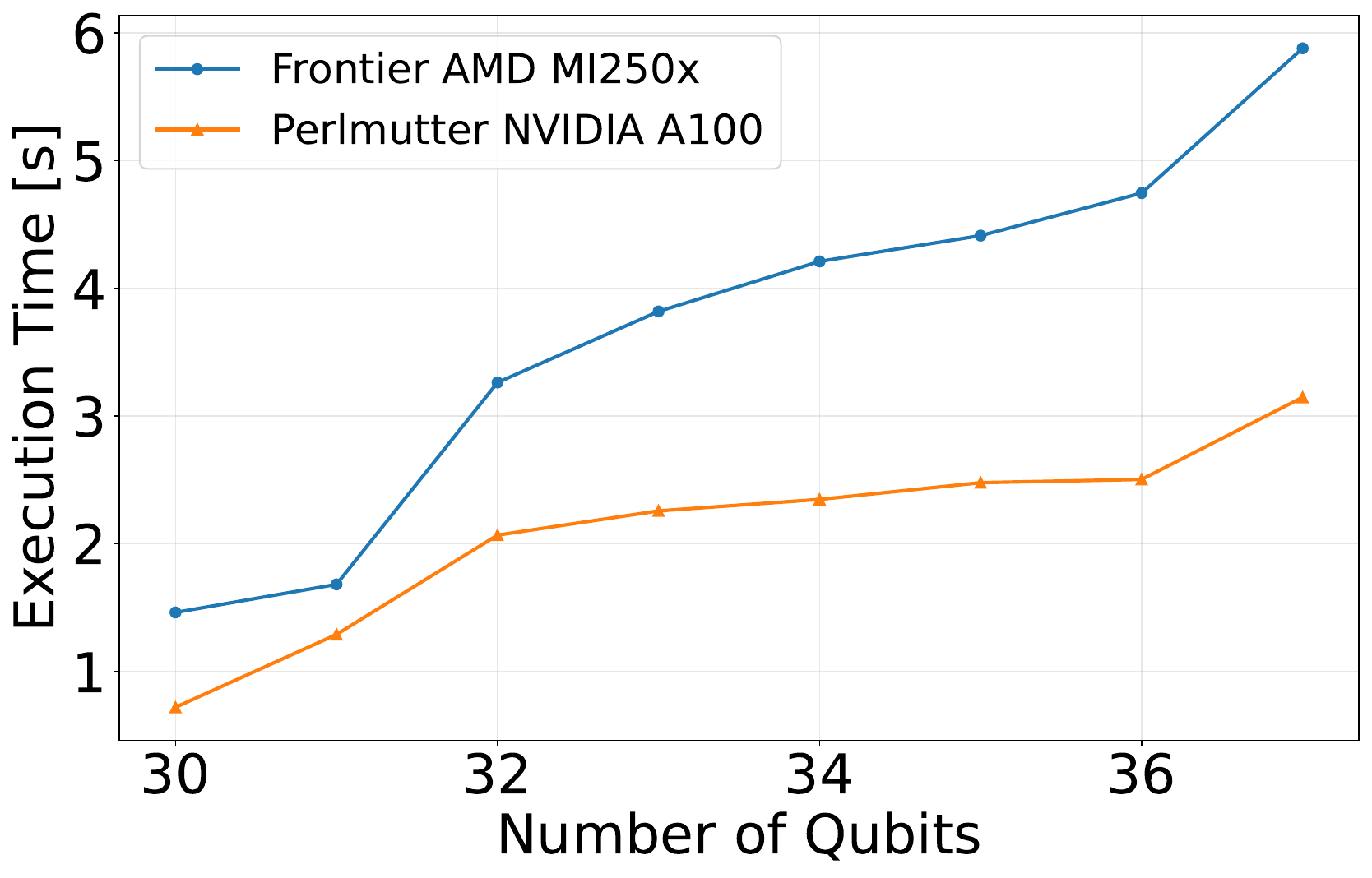}
    \caption*{Ising Algorithm}
\end{minipage}\\[0.5cm]

\begin{minipage}[b]{0.23\textwidth}
    \centering
    \includegraphics[width=\textwidth]{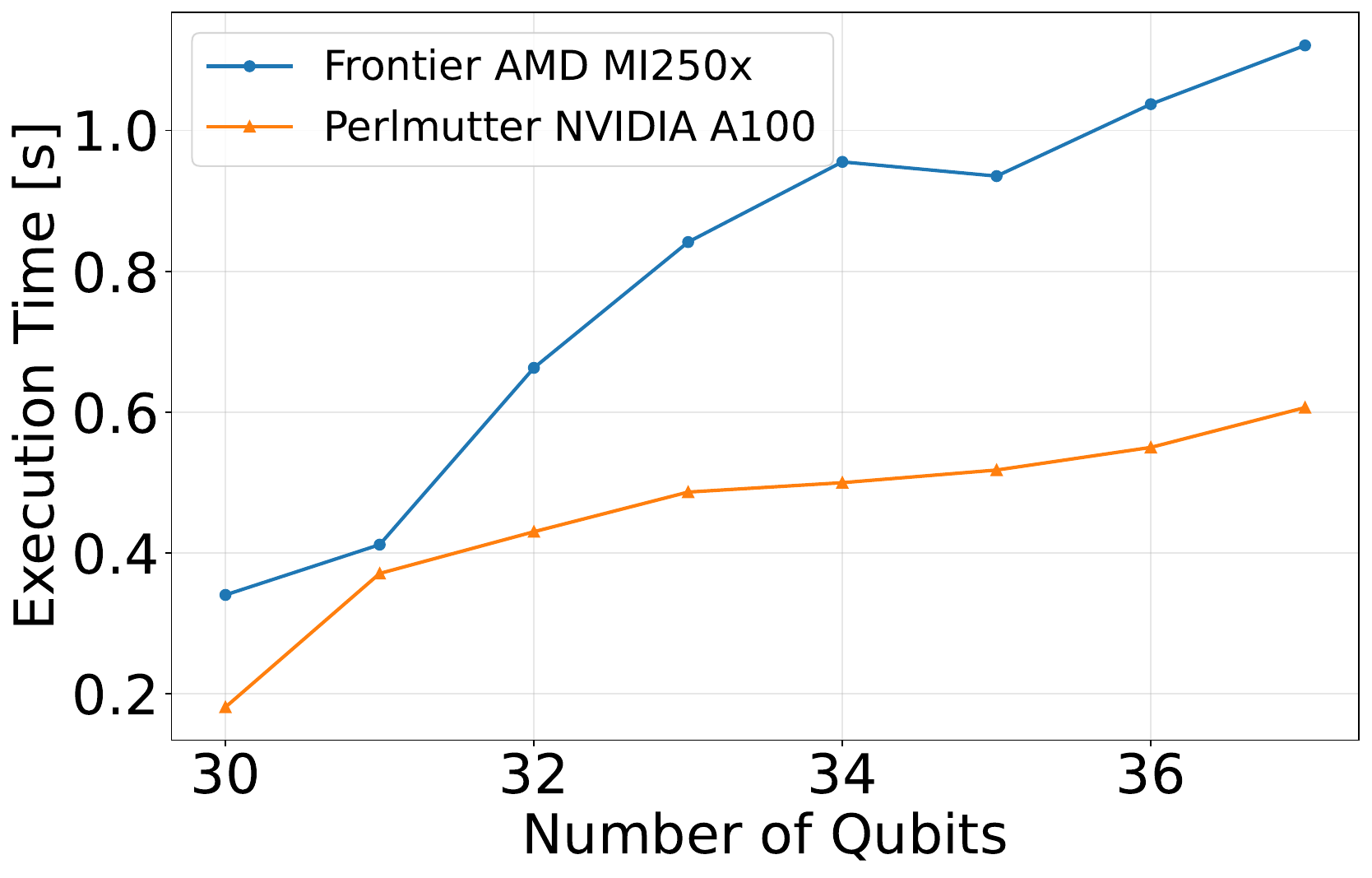}
    \caption*{QFT Algorithm}
\end{minipage}
\hfill
\begin{minipage}[b]{0.23\textwidth}
    \centering
    \includegraphics[width=\textwidth]{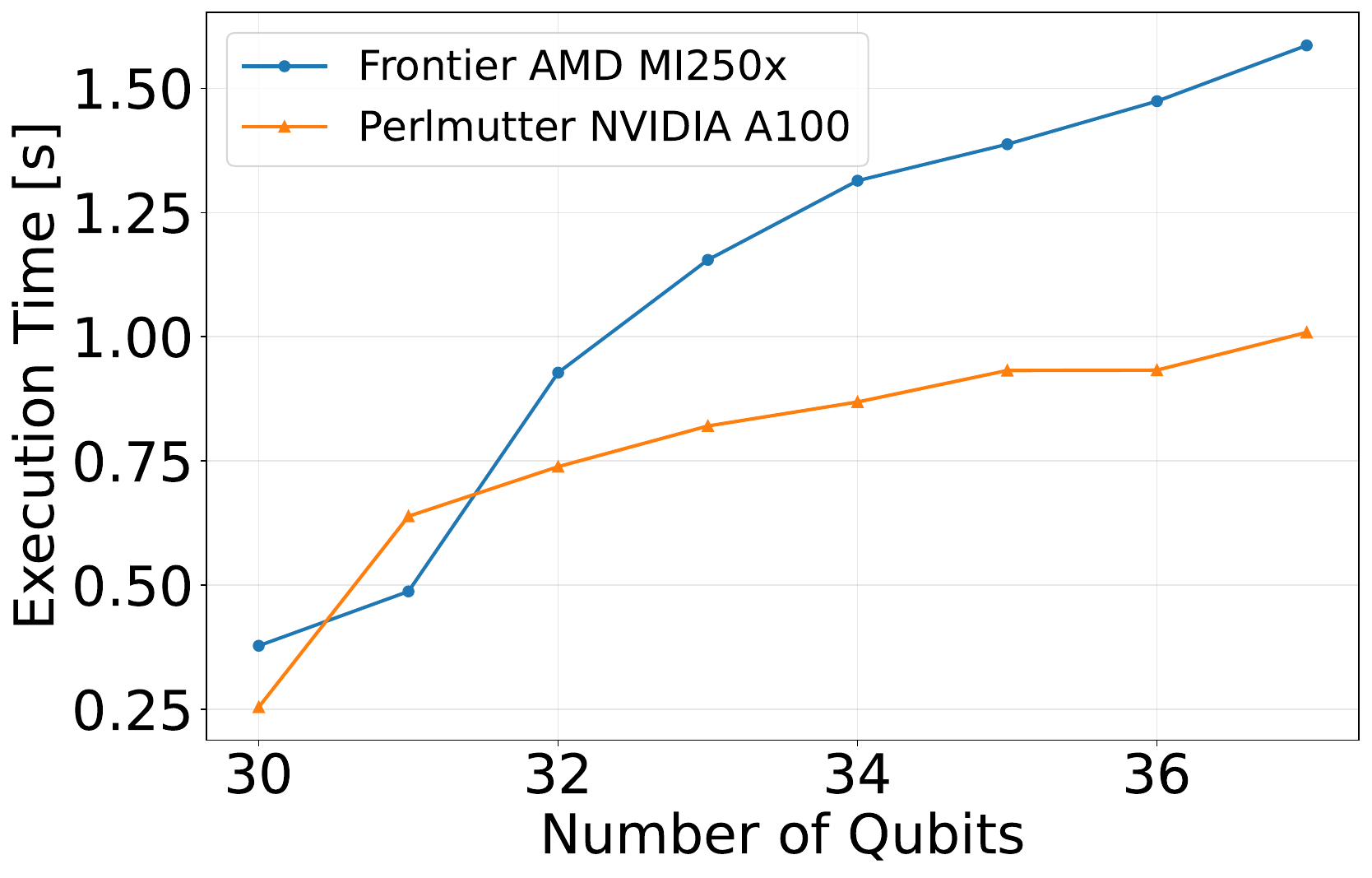}
    \caption*{QPE Exact Algorithm}
\end{minipage}
\hfill
\begin{minipage}[b]{0.23\textwidth}
    \centering
    \includegraphics[width=\textwidth]{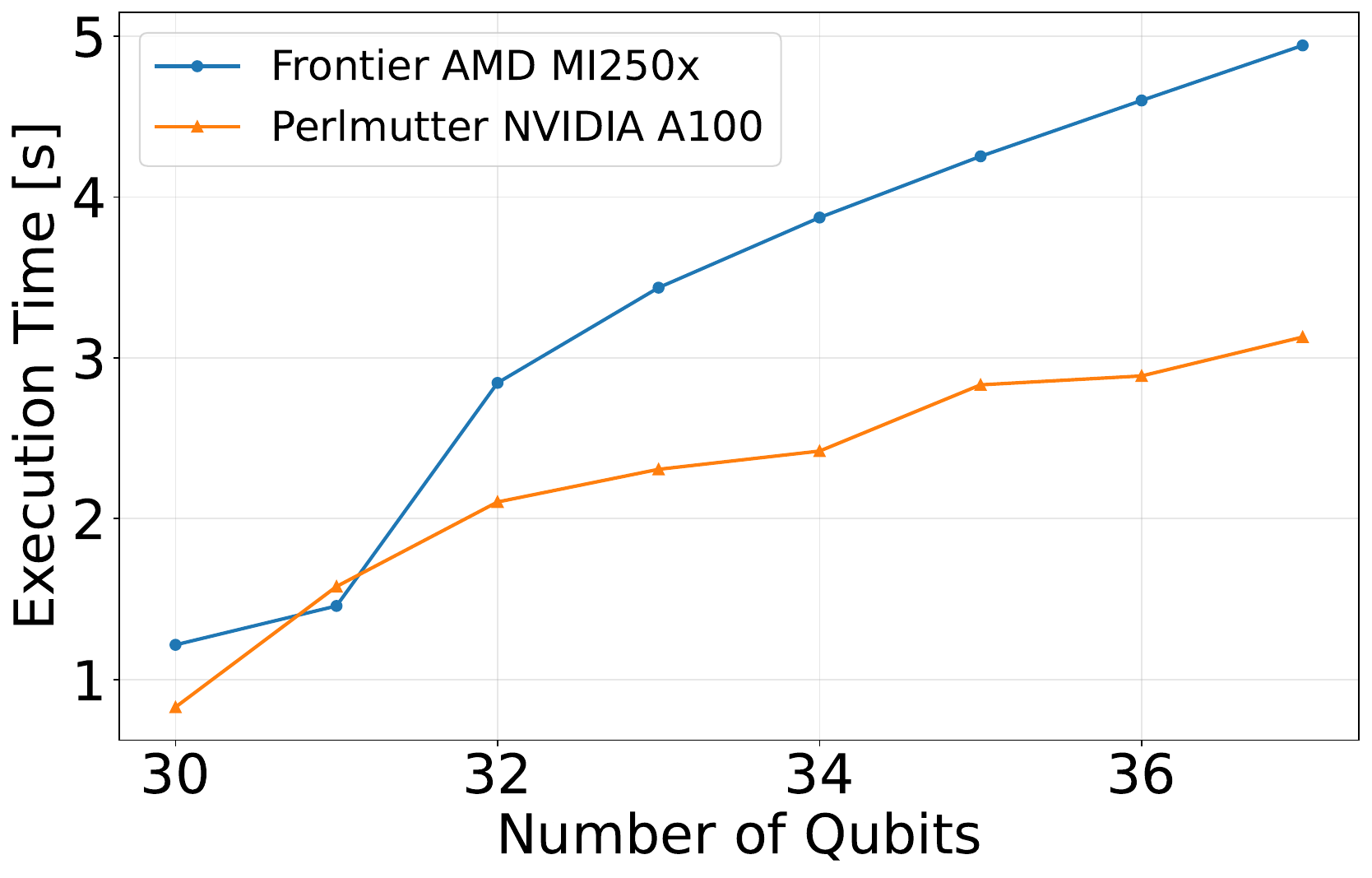}
    \caption*{SU2 Random Algorithm}
\end{minipage}
\hfill
\begin{minipage}[b]{0.23\textwidth}
    \centering
    \includegraphics[width=\textwidth]{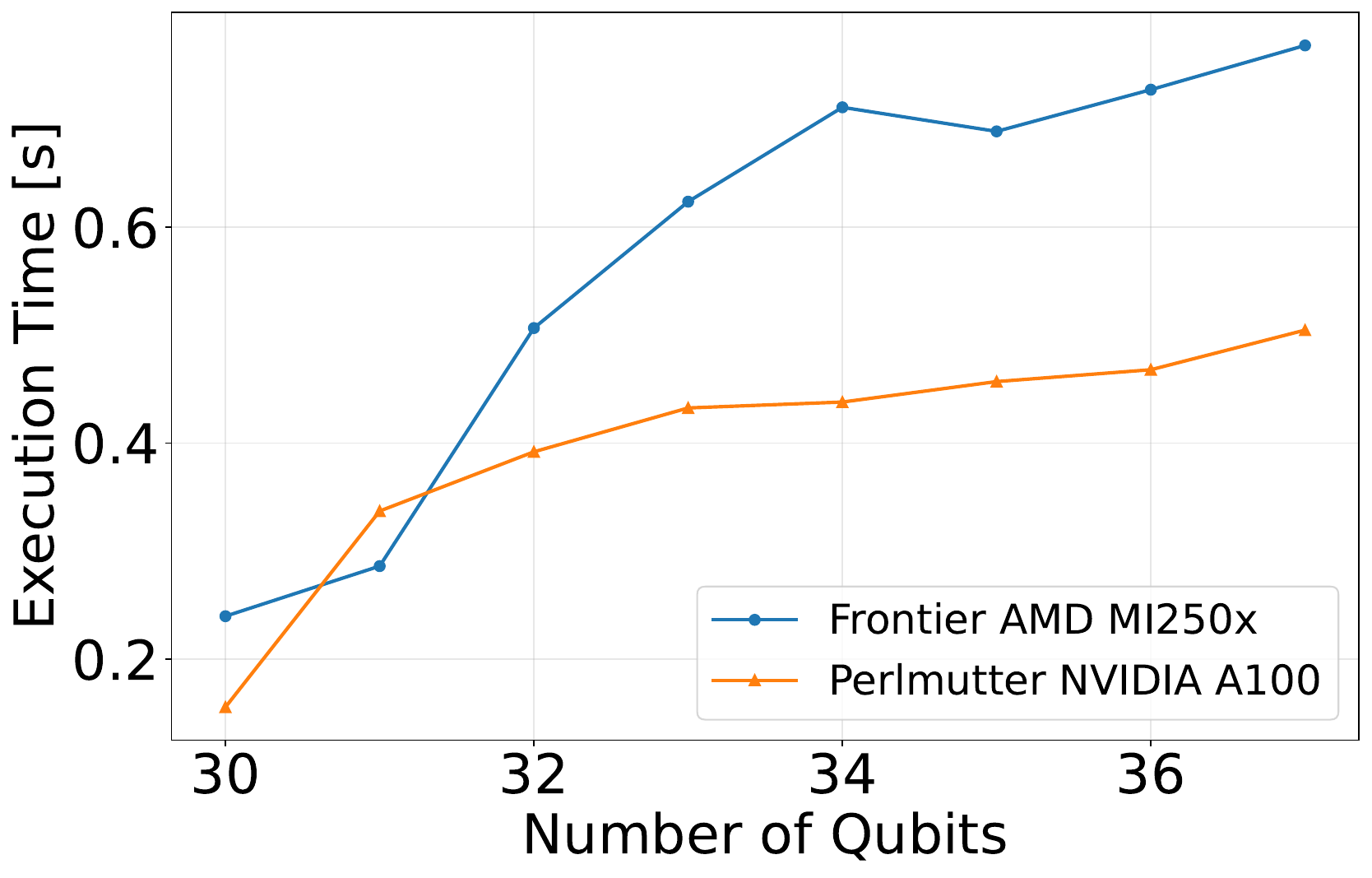}
    \caption*{VQC Algorithm}
\end{minipage}

\caption{Weak scaling results for eight quantum algorithms using our optimized approach for partitioning for both the shared memory and the registers. There is no baseline implementation on AMD GPUs.}
\label{fig:algorithms_amd}
\end{figure*}

\textbf{Comparison with the baselines.} Figure~\ref{fig:all_algorithms} presents weak scaling results for three different approaches on the Perlmutter GPU system. We report execution time in milliseconds. For each experiment, we increase the number of qubits from $30$ to $36$. We start with $4$ GPUs for the $30$ qubit case. As we increase the number of qubits, so does the number of GPUs increases. We report results for Qiskit and cuQuantum, Atlas and our approach using three different optimizations. For our implementation, we experiment with one level of partitioning for distributed memory only ("no local fuse"). Then we use distributed memory and shared memory (two level memory hierarch). Finally, we partition the circuit one more based on a the number of registers. We show the different optimizations to emphasize the key optimizations needed to obtain high performance implementations for a all quantum circuits. Overall, our fully optimized approach delivers clear performance improvements over the Qiskit + cuQuantum. Moreover, our approach is competitive with the Atlas framework. It is worth mentioning, that our implementation parses the circuits as they are and does not remove any gates from the original circuits. Both Qiskit + cuQuantum and Atlas remove \texttt{SWAP}, which means their experiments have less data movement. We are investigating how to modify the inner code of Atlas to allow for all types gates. Moreover, the current experiments report only $36$ qubits. We ran our approach on $37$, however for the baseline implementation, we either could not get the results or for some large circuits the solvers timed out.

Figure~\ref{fig:partition_time} shows the scaling of our partitioning algorithm for a wide range of quantum circuits. All the partitions were executed on the CPU side. For this set of experiments, we report the overall time for the the three level memory hierarchy case (distributed memory, shared memory and local memory). Overall, the our algorithms requires under one second for most of the circuits. For more complicated circuits, like \texttt{qnn} and \texttt{vqc}, the total time to perform the partitioning was approximately $4$ and $5$ seconds, respectively. The implementations could further be improved if the partitioning is re-written using C++ rather than Python. For this set of optimization, we leave it as future work. Overall, the time to find partitions offers a flexible tool that can enable faster exploration of large scale quantum circuits.

\textbf{Extending the computation to the AMD platform.} Figure~\ref{fig:algorithms_amd} presents weak scaling results obtained on the Frontier system using multiple AMD MI250x. We compare the results with the ones obtained by our approach on the NVIDIA platform. For each experiment, we increase number of qubits from $30$ to $37$. Accordingly, we also increase the number of GPUs. Overall, the experimental results are similar across the different quantum circuits. The first few experiments that use $4$ and $8$ GPUs outline that the results on the AMD platform can become competitive with the ones obtained on the NVIDIA platform. However, as the number of GPUs and hence qubits is increased, the overall trends changes. The former is caused by the fact that Frontier has 4 AMD GPUs, each GPU having 2 smaller chips, so in total $8$ GPUs. The Perlmutter system has only $4$ GPUs per node. Communication between the GPUs on the same node is cheaper compared to communication between GPUs on different compute nodes. 

\begin{figure}[t]
    \centering
    \includegraphics[width=0.8\linewidth]{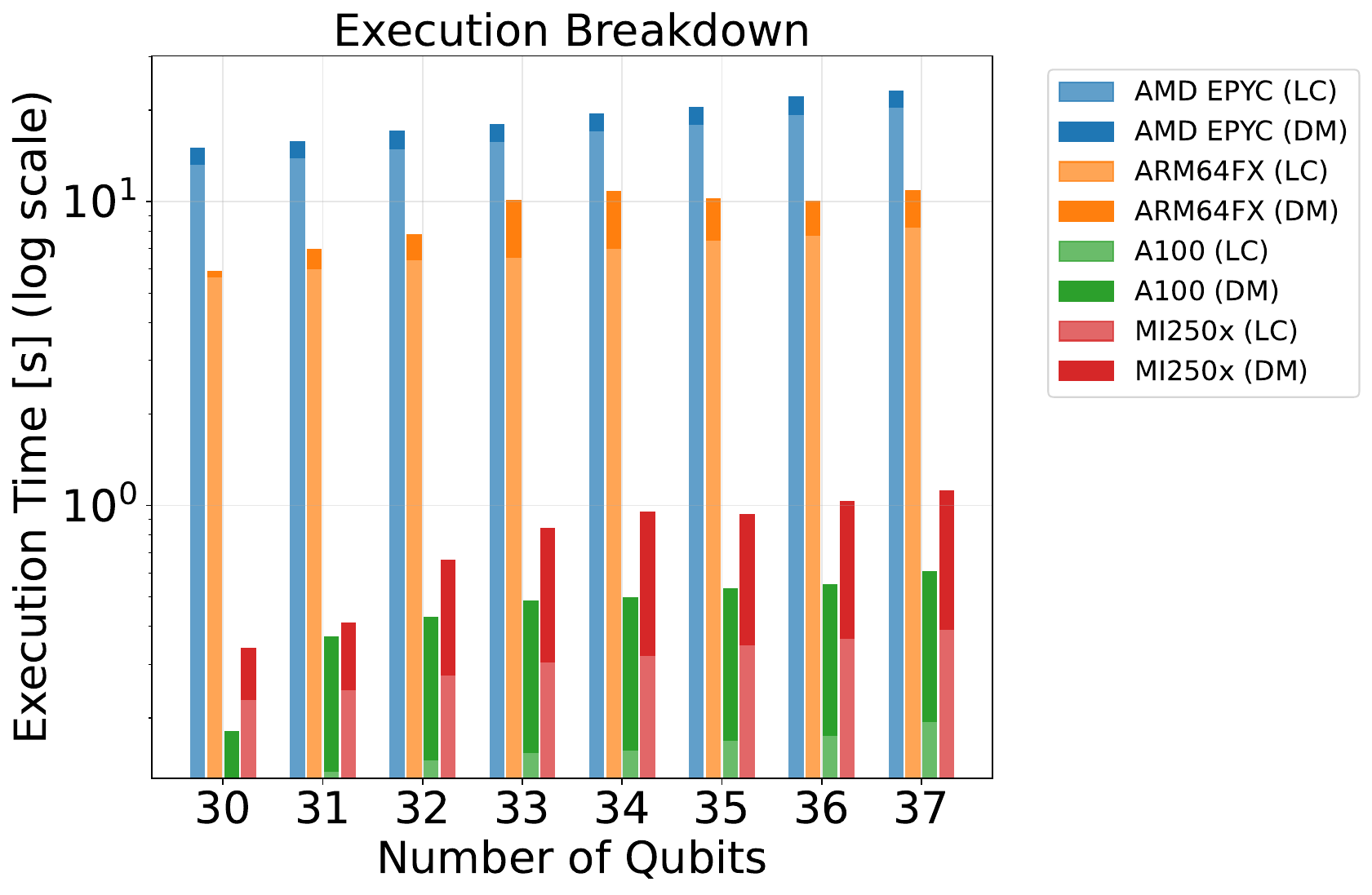}
    \caption{Breakdown of the execution time for the QFT quantum circuit on the four different platforms. The execution time is broken into time spent in local computation (LC - bottom bar), and time spent in moving data between compute nodes (DM - top bar).}
    \label{fig:breakdown}
\end{figure}

\textbf{Execution breakdown.} Figure~\ref{fig:breakdown} shows the breakdown of the execution time only for the QFT quantum circuit on both multi-CPUs and multi-GPUs. For each experiment, we report the execution time for the local computation (LC - bottom) and data movement (DM - top) between the compute nodes as stacked bars. We present the breakdown for the different number of qubits. We show the execution time in log scale, due to the large range between the CPU and GPU results. The first two bars correspond to the results obtained on the CPU systems (AMD EPYC and ARM64FX). The last two bars represent the results for the two GPU systems (NVIDIA A100 and AMD MI250x). For the CPU experiments, the computation dominates the overall execution time. However, when the simulation is ported to the GPU systems, the local computation becomes negligible. Communication will dominate the overall execution time. Therefore, strategies to minimize data movement between the compute nodes is required. Overall, we show that our approach can provide a flexible implementation that can provide support for both CPU systems. While the current implementation offers a competitive and portable solution, there is still work to be done. Specifically, on the CPU side, the generated code focuses on scalar code. Therefore, computation can be improved if SIMD instructions are to be used. Similarly, some of the GPU operations require data reshaping to better utilize the shared memory. Changing the data layouts will provide more efficient local computation. We leave all these features as future work.

\section{Conclusion}
\label{sec:conclusions}
Overall, in this work we have provided an end-to-end solution that translates quantum circuits into efficient and high-performance code for large-scale quantum simulations. We developed this flexible framework by casting quantum circuits and quantum gates as tensors and tensor operations. By capturing quantum circuits as contraction graphs, we developed a novel partitioning algorithm based on closeness centrality that effectively clusters quantum gates according to a system's memory hierarchy while significantly minimizing data movement between compute nodes. Our lightweight code generator successfully translates the obtained partitions into efficient code that runs seamlessly across diverse HPC platforms, from CPU clusters to GPU-accelerated supercomputers. Through comprehensive benchmarking, we demonstrated that the framework offers competitive implementations to state-of-the-art, while offering a flexible solution on more than NVIDIA platforms. This work establishes a foundation for more accessible and efficient quantum circuit simulation, bridging the gap between theoretical quantum algorithms and practical computational validation on current HPC infrastructure.

\bibliographystyle{ACM-Reference-Format}
\bibliography{biblio}

\end{document}